\documentclass[pdf,12pt]{amsart}
\usepackage{times}
\usepackage{curves}
\usepackage{amssymb,amsfonts,eufrak,amsmath,amscd}
\usepackage{amsmath}
\usepackage[pdftex]{graphicx}
\begin{document}
\newtheorem{theo}{Theorem}
\newtheorem{lem}{Lemma}
\newtheorem{cor}{Corollary}
\newtheorem{prop}{Proposition}
\newtheorem{ex}{Example}
\newtheorem{remar}{Remark}
\newtheorem{rul}{Rule}
\newtheorem{conj}{Conjecture}
\newtheorem{defi}{Definition}

\newcommand{\bt}{\begin{theorem}\em}
\newcommand{\et}{\end{theorem}}
\newcommand{\petet}[2]{{}^{#2}\!{#1}}
\newcommand{\side}[2]{{}^{#1}\!{#2}}

\newcommand{\mop}[2]{\stackrel{{}_{{#2}}}{#1}}
\newcommand{\Mop}[4]{\overset{\ \ \ #4}
{\underset{\ \ \ #3}
{\mathop{\sideset{^{{}_{{#2}}}}{}#1}}}}
\newcommand{\la}{\langle}
\newcommand{\ra}{\rangle}
\newcommand{\ovl}[1]{\overline{#1}}
\newcommand{\Bak}[1]{\mathbf B_{#1}}
\newcommand{\bak}{\mathbf b}
\newcommand{\dg}[2]{\mathrm{ex}_{#1}#2}
\newcommand{\ord}[2]{\mathrm{ord}_{#1}#2}
\newcommand{\Ex}[1]{\!\!\uparrow^{#1}}
\newcommand{\Exx}[2]{(#1)\!\!\uparrow^{#2}}
\newcommand{\STDm}[1]{\mathrm{STD}({#1})}
\newcommand{\STD}[1]{{\rm STD}$({#1})$}
\newcommand{\supp}{\mathrm{supp}} 
\newcommand{\s}[1]{{#1}^{\star}}

\newcommand{\beas}{\begin{eqnarray*}}
\newcommand{\eeas}{\end{eqnarray*}}
\newcommand{\bea}{\begin{eqnarray}}
\newcommand{\eea}{\end{eqnarray}}
\newcommand{\lcm}{\mathrm{lcm}}
\newcommand{\md}{\mathrm{mod\ }}
\newcommand{\expon}[2]{\mathrm{ex}_{#1}\left({#2}\right)}
\newcommand{\z}{\mathbf 0}
\newcommand{\rem}{\mathrm{rem}}
\newcommand{\para}{\parallel}
\newcommand{\rt}[1]{\mathrm{rem}\left(#1,\tau\right)}
\newcommand{\mulmi}[1]{\left\lfloor #1\right\rfloor_{\tau}}
\newcommand{\mulma}[1]{\left\lceil #1\right\rceil_{\tau}}
\newcommand{\mm}[1]{\,(\mathrm{mod}\ #1)}
\newcommand{\mv}[1]{\, \mathrm{mod}(#1)}

\newcommand{\pr}{{\sl Proof.\ }}
\newcommand{\bx}{\ \ \ $\Box$}
\newcommand{\bxm}{\ \ \ \Box}
\newcommand{\ind}{\mathrm{ind}\,}
\newcommand{\Acal}{\mathcal{A}}
\newcommand{\Sal}{\mathcal{S}}
\newcommand{\Pal}{\mathcal{P}}
\newcommand{\Hal}{\mathcal{H}}
\newcommand{\Mal}{\mathcal{M}}
\newcommand{\Ral}{\mathcal{R}}
\newcommand{\Cal}{\mathcal{C}}
\newcommand{\ccirci}[2]{\mathcal{C}(#1)\big|_{#2}}
\newcommand{\rcirci}[2]{\mathcal{R}(#1)\big|^{#2}}
\newcommand{\Bfrak}{\mathfrak B}
\newcommand{\Bbol}{\mathbf B}

\newcommand{\mb}[1]{\mathbb{#1}}
\newcommand{\m}[1]{\mathfrak{#1}}
\newcommand{\bfs}{\text{ \bf: }}
\newcommand{\mbf}[1]{\mathbf{#1}}
\newcommand{\proj}{\mathrm{proj}}
\newcommand{\ebxt}[2]{{#1}^{\boxtimes#2}}
\newcommand{\bxt}[2]{{#1}\boxtimes{#2}}
\newcommand{\BT}{\boxtimes}
\newcommand{\setp}{\{0,\dots,p-1\}}
\newcommand{\aut}[1]{\mathbf {Aut}(#1)}
\newcommand{\debt}[2]{{#1}^{\BT{#2}}}
\newcommand{\cst}{\circledast}
\newcommand{\ED}{\end{document}}
\newcommand{\mind}[1]{\widetilde{#1}}

\sloppy

\title[Reduction]{Reduction of behavior of additive cellular automata\\ on groups}

\author[V. Bulitko]{Valeriy Bulitko\\
\ \\
{\tiny Athabasca University}\\
{\tiny valeriyb@athabascau.ca}}

\vspace*{-2.5cm} \sloppy \thispagestyle{empty}

\DeclareGraphicsExtensions{.png,.gif,.jpg,.pdf}

\maketitle

\vspace{-0.5cm}

\begin{abstract}
A class of additive cellular automata (ACA) on a finite group is defined by an index-group $\m g$ and a finite field $\m F_p$ for a prime modulus $p$ \cite{Bul_arch_1}. This paper deals mainly with ACA on infinite commutative groups and direct products of them with some non commutative $p$-groups. 
It appears that for all abelian groups, the rules and initial states with finite supports define behaviors which being restricted to some infinite regular series of time moments become significantly simplified. In particular, for free abelian groups with $n$ generators states $V^{[t]}$ of ACA with a rule $R$ at time moments $t=p^k,k>k_0,$ can be viewed as $||R||$ copies of initial state $V^{[0]}$ moving through an $n$-dimensional Euclidean space. That is the behavior is similar to  gliders from J.Conway's automaton {\sl Life}. For some other special infinite series of time moments the automata states approximate  self-similar structures and the approximation becomes better with time. An infinite class $\mathrm{DHC}(\mbf S,\theta)$ of non-commutative $p$-groups  is described which in particular includes quaternion and dihedral $p$-groups. It is shown that the simplification of behaviors takes place as well for direct products of  non-commutative groups from the class $\mathrm{DHC}(\mbf S,\theta)$ with commutative groups. Finally, an automaton on a non-commutative group is constructed such that its behavior at time moments $2^k,k\ge2,$ is similar to a glider gun. It is
concluded that ACA on non-commutative groups demonstrate more diverse variety of behaviors comparing to ACA on commutative groups.
\end{abstract}\ 

\indent{\bf Subj-class:} nlin. CG\\
\indent{\bf MSC-class:} 37B15, 68Q80

\vspace{-0.1 cm}

\footnotesize
\tableofcontents

\normalsize

Lt $\m g,\m F_p$ be a group (index-group) and the finite field by modulo $p$. An element $g\in\m g$ given, let $g^{-1}$ and $\ovl g$ denote the inverse element for $g$ in $\m g$, i.e. $g\ovl g=\ovl gg=\m 1$ where $\m1$ (or $\m 1_{\m g}$) is the unit of $\m g$. 

By $v(g)$ we denote $g$-th component of a string $v$ where $g\in\m g$. Often it is also convenient by $v(g)$ to denote $v\big|^g$. A pair 
$(\m g,\m F_p)$ defines the semigroup $\m M[\m g,\m F_p]$ of strings $\{V|V(g)\in\m F_p,g\in\m g\}$ with multiplication $\BT$ that is a group convolution (see references in~\cite{Bul_arch_1})
\bea\label{BT}
(V\BT W)\Big|^k=\sum_{s\in\m g}V(s)W(\ovl sk),k\in\m g.
\eea
Here $U|^k$ for $U\in\m M,k\in\m g$ is another convenient notation for $U(k)$. All operations with elements of $\m F_p$ are performed in the field, and operations with group elements - in the group. Rules and states are elements of $\m M$. We use common name {\sl vectors} for elements of $\m M[\m g,\m F_p]$.

Let $V$ be a current state of an additive cellular automaton (ACA) with a rule $R$. 

{\sl Application} $R*V$ of rule $R$ to state $V$ yields next state $V'$ of the automaton. The next state $V'$ of the automaton can be computed as $V'=V\BT R^-$ where $(R^-)\big|^i=R(-i),i\in\m g$ (see \cite{Bul_arch_1}).\footnote{We distinguish the denotation ``$\,\ovl g$\,'' used for reverse elements of index groups from ``$\,v^-$'' denoting elements of certain kind in semi-groups.}

Let $\supp(A)$ denote the set $\{g|A(g) \neq 0\}$ for a vector $A$. we say that vectors $A,B$ are {\sl disjoint} if $\supp(A)\cap\supp (B)=\emptyset$.

Let $\m M^f=\{V\in\m M\bfs |\supp(V)| <\infty\}$. We call elements of $\m M^f$ and any automaton whose initial state and rule belong to $\m M^f$ {\sl finite}.

If a group $\m g$ commutative, by $\m 0$ and $+$ we denote the group unit and the group operation. As usual $\underbrace{g+\dots+g}_{k\text{ terms}}=k\cdot g, g\in\m g.$

$\mathbb I$ is the unit of semigroup $\m M[\m g,\m F_p]$. By $1,0$ we denote the unit and zero of the field $\m F_p$. Also by $\mathbb O$ denote zero vector in $\m M[\m g,\m F_p]$. 
$V\in\m M[\m g,\m F_p]$ and $r\in\m F_p$ given, the vector $rV$ is defined by the condition $(rV)|^g=r(V|^g),g\in\m g$.

\section{Convolution on groups}

Here are some elementary properties of $\BT$ (some properties with  references see in \cite{Bul_arch_1}).
Let $\m K$ be the commutant of an index group $\m g$. We say that $A\in\m V$ is $\m K$-{\sl correct} if  $\forall g,f[g\ovl f\in\m K\implies A(g)=A(f)]$. 

An element $g\in\m g$ given, the vector $K[g]$ is defined as 
\beas
K[g](h)=
\begin{cases}
 1,& \text{ if } h=g\\
0,& \text{ else.}
\end{cases}
\eeas 
Obviously we have $A=\sum_{g\in\supp(A)}A(g)K[g]$ for any $A\in\m M[\m g,\m F_p]$.

\begin{lem}\label{oper-bxt}
\begin{itemize}
\item[(i)] $\bxt{}{}$ is an associative operation.\\
\item[(ii)] $K[f]\BT K[g]=K[fg]$ for any $f,g\in\m g$.\\
\item[(iii)] 
$(K[g]\BT A)(f)= A(\ovl gf),\ \ \ \ (A\BT K[g])(f)= A(f\ovl g)$.\\
In particular, $\bxt{K[\m1]}A=\bxt A{K[\m1]}=A$.\\
\item[(iv)] 
$
A\BT(B+C)=A\BT B+A\BT C,\ \ (B+C)\BT A=B\BT A+C\BT A.
$\\
\item[(v)] $A^{-}\BT B^{-}=(B\BT A)^{-}$.
\item[(vi)]   $A\BT B=B\BT A$ for any $A,B\in\m V$ such that at least one of them is $\m K$-correct.\\
\item[(vii)] If $G$ is commutative group then $\bxt AB=\bxt BA$.\\
\item[(viii)] Let $H$ be a subgroup of $\m g$ and $\supp(A), \supp(B)\subseteq H$. Then $\supp(\bxt AB)\subseteq H$. In addition if $H$ is commutative then $A\BT B=B\BT A$. 
\end{itemize}
\end{lem}
\pr (i) For $A,B,C\in\m V,g\in\m g$ we have
\beas
[\bxt{(\bxt AB)}C](g)=\sum_f\left(\sum_rA(r)B(\ovl rf)\right)C(\ovl fg)= \sum_rA(r)\sum_fB(\ovl rf)C(\ovl fg)=_{h:=\ovl rf} \\ =\sum_rA(r)\left(\sum_{h}B(h)C(\ovl h\ovl rg)\right)= \sum_rA(r)(\bxt BC)(\ovl rg)=[\bxt A(\bxt BC](g).
\eeas
It is possible to replace bounded variable $f$ with $h$ since for each fixed value of $r$ mapping $f\mapsto \ovl rf$ is 1-1-mapping $\m g$ on $\m g$.\footnote{We use similar arguments below several times.}\\
(ii)
\beas
(K[f]\BT K[g])(r)=\sum_sK[f](s) K[g](\ovl sr) =K[g](\ovl fr)=
\begin{cases}
1,& \ovl fr=g\\
0& \text{otherwise}.
\end{cases}
\eeas
Since $\ovl fr=g$ means $r=fg$ we get what we need.\\
(iii)
\beas
(K[g]\BT A)(f)=\sum_sK[g](s)A(\ovl sf)=_{\text{since }K[g]=0\text{ if }g\neq s}=A(\ovl gf).\\
(A\BT K[g])(f)=\sum_sA(s)K[g](\ovl sf)=_{\text{since }K[g]=0\text{ if }g\neq \ovl sf}=A(f\ovl g).
\eeas
In particular,
$(\bxt{K[\m1]}A)(g)=(\bxt{A}{K[\m1]})(g)=A(g)$ that is $\bxt{K[\m1]}A=\bxt A{K[\m1]}=A$\\
(iv) Now
\beas
(A\BT(B+C))(g)=\sum_fA(f)(B+C)(g\ovl f)=\sum_fA(f)B(\ovl fg)+ \sum_fA(f)C(\ovl fg)=\\(A\BT B)(g)+(A\BT C)(g).
\eeas
The second identity has a similar proof.\\
(v) $(R^{-}\BT Q^{-})(g)=\sum_{f}R^{-}(f)Q^{-}(\ovl fg)=\sum_{f}R(\ovl f)Q(\ovl gf)$. If we define $w=\ovl gf$ (for any fixed $g$ variable $w$ runs over $\m g$ while $f$ runs over $\m g$) then $\ovl f=\ovl w\,\ovl g$ and $\sum_{f}R(\ovl f)Q(\ovl gf)=\sum_{w}Q(w)R(\ovl w\,\ovl g)=(Q\BT R)(\ovl g)=(Q\BT R)^{-}(g)$. \bx\\ 
(vi) By definition we can write:
\beas
(\bxt AB)(g)=\sum_fA(f)B(\ovl fg)=\sum_rB(r)A(g\ovl r)\\
(\bxt BA)(g)=\sum_rB(r)A(\ovl rg)=\sum_fA(f)B(g\ovl f).
\eeas
Now if for instance $B$ is $\m K$-correct then $B(\ovl fg)=B(g\ovl f)$ for all $g,f\in\m g$. This is because $g\ovl f\,\ovl{\ovl fg}$ is a commutator. Hence $(\bxt AB)(g)=(\bxt BA)(g)$.\\
(vii) This is the prompt consequence of (vii) due to the fact that for commutative group $\m K=\{\m1\}$ in which case any vector is $\m K$-correct. \\
(viii) By definition we can write:
\beas
(\bxt AB)(g)=\sum_{f\in G}A(f)B(\ovl fg)=
\sum_{f\in H}A(f)B(\ovl fg)
\eeas
where if $B(\ovl fg)\neq0$ then $\ovl fg\in H$. With $f\in H$ this means $g$ must belong to $H$ in case when $(\bxt AB)(g)\neq0$.

Assume $H$ is a commutative subgroup of $\m g$. In this case 
\beas
(\bxt AB)(g)=\sum_{f\in \m g}A(f)B(\ovl fg)=
\sum_{f\in H}A(f)B(\ovl fg)=_{\text{due to }\ovl fg=g\ovl f}
\sum_{f\in H}A(f)B(g\ovl f)=_{w:=g\ovl f}\\ 
\sum_{w\in H}A(\ovl wg)B(w)=\sum_{w\in \m g}B(w)A(\ovl wg)=(B\BT A)(g). 
\eeas
\bx

\begin{cor}\label{cor_for-} 
If $A^-=A$ then for any $n\in\mathbb Z^+$ we have  
$(A^{\BT n})^-=A^{\BT n}$.
\end{cor}

Let $\m h$ be a subgroup of $\m g$. 
We call vector $A$ {\sl right-$[a,\m h]$-symmetric} if 
\begin{enumerate}
\item $\forall g[A(g)\neq0\implies \exists q[g=aq]]$;
\item $\forall g[A(ag)=A(ag^{-1})]$.
\end{enumerate}

\section{Baker transformation for additive CA on group $\m g$}

\subsection{Case of commutative $\m g$}

Let us use notation  $\debt Tn$ for $\underbrace{T\BT\dots\BT T}_{n\text{ terms}}$ as $T\in\m M$ and define $\bak_{p}(T)$ by the condition
\bea\label{2}
\bak_p(T)|^k=
\begin{cases}
\phantom{nbbb}0,&\text{ if  }\{g\in\m g|pg=k\}=\emptyset,\\
\underset{i\in\{g\in\m g|p\cdot g=k\}}{\sum}T(i),&\text{ otherwise}.
\end{cases}
\eea
\begin{lem}
\beas
(T_1\BT\dots\BT T_n)\Big|^k=\underset{i_1,\dots,i_{n-1}\in\m g}{\sum} T_1(i_1)\cdot\dots\cdot T_{n-1}(i_{n-1})T_n(k-i_1{}_{\dots}-i_{n-1})
\eeas
\end{lem}
\pr 
\bx

We call a group $\m g$ {\sl $p$-normal} if the set $\{g\in\m g| p\cdot g=\m0\}$ is finite.\footnote{That is the number of elements of order $p$ is finite.} Let $\kappa=|\m p|$. It is also the number of terms in sums from definition (\ref{2}). Clearly for any $q\in\m g$
\bea
\exists r\in\m g[p\cdot r=q]\implies |\{g\in\m g|p\cdot g=q\}|=\kappa.
\eea
This is because $\{g\in\m g|p\cdot g=\m0\}$ is a 
subgroup of $\m g$. 

\begin{theo}\label{baker-commut}
(i) $\bak_p$ is well defined on $\m M^f$. If $\m g$ is $p$-normal then $\bak_p$ is well defined on $\m M$. \\
(ii) $\forall T\in\m M[\debt Tp=\bak_p(T)]$ for any prime $p$ and any abelian $p$-normal index-group $\m g$.
\end{theo}
\pr (i) It obvious. 
(ii) Since
\bea\label{theo1-proof}
\debt Tp \Big|^k=\underset{i_1,\dots,i_{p-1}\in \m g}{\sum}T(i_1)\cdot\dots\cdot T(i_{p-1}) T(k-i_1\dots-i_{p-1})
\eea
We partition the collection of $T(i_1),\dots,T(i_{p-1}),T(k-i_1\dots-i_{p-1})$ on subsets consisting of elements $T(i_k)$ with equal indices $i_k$. Let there be $r$ subsets whose numbers of elements
are $n_1,\dots,n_r$ where $n_1+\dots+n_r=p$. Then we use the fact that $multinomial(n_1,\dots,n_r)$ is multiple of prime $p$ for cases when  $r>1$ and $n_j>0, j=1,\dots,r$.
The case\footnote{Recall that operations $+,\cdot$ are defined in $\m F_p$ by modulo $p$} when  $i_1=i_2=\dots=i_{p-1}=k-i_1-\dots i_{p-1}$ only remains. Thus
\beas
\debt Tp \Big|^k=\underset{i\in\{j\in\m g|p\cdot j=k\}}{\sum}(T(i))^p
\eeas
On the other hand $f^p=f$ for any $f\in\m F_p$.
\bx

We use powers of $\bak_p$ according to the definition 
\bea\label{b^k}
\bak_p^0(V)=V,\quad \bak_p^{k+1}(V)=\bak_p(\bak_p^k(V)).
\eea 

\subsection{$\m g$ is not commutative}

In general the theorem~\ref{baker-commut} is not true for non commutative groups. However it works for some special classes of rules. 
\begin{theo}\label{baker-noncommut}
Let $\m g$ be any $p$-normal\footnote{This condition is not necessary if we restrict $\bak_p$ to $\m M^f$.} group, $\m K$ its commutant, and $\m a$ a commutative sugroup of $\m g$. If $T$ is $\m K$-correct or $\supp(T)\subseteq\m a$ then $\debt Tp=\bak_p(T)$.
\end{theo}
\pr  
Assume that $T$ is $\m K$-correct. Now we use multiplicative notation for products of group elements and rewrite~(\ref{theo1-proof}) as follows
\beas
\debt Tp \Big|^k=\underset{i_1,\dots,i_{p-1}\in \m g}{\sum}T(i_1)\cdot\dots\cdot T(i_{p-1}) T(\ovl{i_1\dots i_{p-1}}\,k).
\eeas 
Due to the condition that $T$ is $\m K$-correct, the value of
$T(\ovl{i_1\dots i_{p-1}}\,k)$ does not depend on order of the group elements in the product  $\ovl{i_{p-1}}\dots\ovl{i_1}\,k$: all permutations leave the product in the same adjacent class of $\m g$ respectively $\m K$. Therefore the reasoning from the proof of (ii) from theorem~\ref{baker-commut} is applicable here as well.

For the case when $\supp(T)\subseteq\m a$ for any $i\notin \m a$ it holds $T(i)=0$ and thereby the sum 
\beas
\underset{i_1,\dots,i_{p-1}\in \m g}{\sum}T(i_1)\cdot\dots\cdot T(i_{p-1}) T(\ovl{i_1\dots i_{p-1}}\,k)
\eeas 
obviously could be restricted by elements of the commutative  subgroup $\m a$:
\beas
\debt Tp \Big|^k=\underset{i_1,\dots,i_{p-1}\in \m a}{\sum}T(i_1)\cdot\dots\cdot T(i_{p-1}) T(\ovl{i_1\dots i_{p-1}}\,k).
\eeas 
Therefore the reasoning from theorem~\ref{baker-commut} works here as well.
\bx

For simplicity below we often use $T^k$ instead of $\debt Tk$.

\section{Decomposition effects for ACA on commutative groups}

\subsection{Case of finite commutative group}\ \\
Decompositional effects for finite commutative groups were described in \cite{Bul_V00r-Bul}. 
Here we reformulate the results. 
Let a cyclic group $C_{n}$ of order $n$ be a direct factor of an index group $\m g$, that is $\m g=G\times C_{n}$ for a group $G$. Assume that $c$ is a generator for $C_n$. If non-negative integer $h$ is the maximal number such that $p^h\mid n$, then there are $p^h$ adjacent classes $i\cdot c+H,i=0,\dots,p^h-1,$ in $C_n$ where $H$ is the subgroup of $C_n$ consisting of $q,q=\frac n{p^h},$ elements $\{{jp^h}\cdot c|j=0,\dots,q-1\}$. Any state $M\in\m M[\m g,\m F_p]$ (we call it {\sl global}) can be represented as a sum  
\beas
\sum_{i=0}^{p^h-1}M_i\quad \text{ of $H$-{\sl adjacent} vectors $M_i$ such that}\quad
M_i(g)=
\begin{cases}
M(g),& \text{if } g\in G\times (ic+H),\\
0,& \text{otherwise}.
\end{cases}
\eeas
The next theorem is about decomposition of behaviors of automata for special time moments. 
\begin{theo} 
A rule $R$ and an initial state $M, (R,M\in\m M[G\times C_n,\m F_p])$ given, by $\tilde R$ denote the vector $\bak_p^{h}(R)$ and by $M_i^{[t]}$ - adjacent state for global state $M^{[t]}$ at time moment $t$. Then
\begin{itemize}
\item $\supp(\tilde R)\subseteq G\times H$;
\item  for any $t\ge1$ and $i\in[0,p^h-1]$ state $M_i^{[t+p^h]}$ is obtained from $M_i^{[t]}$ by application of rule $\tilde R$;
\item for time moments $t=jp^h,j=1,2,\dots,$ evolution of the global state $M$  is assembled as 
\beas
M^{[t]}=\sum_{i=0}^{p^h-1}M_i^{[t]},
\eeas
from the evolutions of adjacent states which are disjoint and could be viewed as evolving independently under action of the same (for all $i$) rule $\tilde R$.
\end{itemize}
\end{theo}
\pr First we have $R^{p^h}=\bak_p^h(R)=\tilde R$. Then,  $R*(M_i+M_j)=(M_i+M_j)\BT R^{-}=M_i\BT R^-+M_j\BT R^-$ (lemma~\ref{oper-bxt}(iv)). Further, assume $(l,k)\in G\times C_n$. Then $(M_i\BT\tilde R^-)|^{(l,k)}=\sum_{(g,r)}M_i(g,r)\tilde R^-(l-g,k-r)\neq0$ only when $\tilde R(l-g,r-k)\neq0$ for some $(g,r)\in G\times C_n$. Therefore $r-k=(r'-k')c$ 
where $r'=k'\mm{p^h}$. 
On the other hand, by definition of $M_i$ to have $M_i(g,r)\neq0$ it must hold that $r\in ic+H$, i.e. $r'=i\mm{p^h}$. From here $k\in ic+H$.
This means that $\tilde R$ transforms vectors whose supports are subsets of some adjacent class of $G\times C_n$ respectively the subgroup $(\m1_G,H)$ into vectors obeying the same condition. 
 \bx

Of course, if the factor group $G$ of the product $G\times C_n$ has itself a direct factor $C_m$ with $m$ divisible by a number $p^{h'},h'>0,$ then at least at moments of time multiple to $p^{max\{h,h'\}}$ the decomposition becomes more deep, so on. 
\begin{cor}
If an index group $\m g$ is a finite commutative group of order $p^h$ then for an automaton with a rule $R$ and an initial state $M$ it holds 
\begin{itemize}
\item $\bak_p^h(R)\in\{f\mathbb I|f=\ovl{0,p-1}\}$;
\item if $\bak_p^h(R)=f\mathbb I$ then for all $t=kp^h$, $k=1,2,\dots$ it holds $M^{[t]}=f^kM$;\footnote{Recall that operations $+,\cdot$ are defined in $\m F_p$ by modulo $p$.}
\item if $\bak_p^h(R)=0\mathbb I$ then for all $t\ge p^h$ we have $M^{[t]}=\mathbb O$.  
\end{itemize}
\end{cor}

\subsection{Case of finitely generated free commutative groups. Gliders and iterative structures}\ \\

We can represent any element $g$ of a free commutative group $\m f_n$ with $n$ generators $g_i,i=\ovl{1,n},$ as $d_1\cdot g_1+\dots+d_n\cdot g_n$
where all $d_i,i=\ovl{1,n},$ are integers i.e. as a point $\chi(g)$ with integer coordinates $(d_1,\dots,d_n)$ in $n$-dimensional Euclidean space $\mathbb E_n$. 
Respectively for any vector $V\in\m M[\m f_n,\m F_p]$ its support has the representation in $\mathbb E_n$ as a set $\chi(\supp(V))=\{\chi(g)|g\in\supp(V)\}$. 

Thus any $V\in\m M[\m f_n,\m F_p]$ can geometrically
represented by a function $\mathcal V:\chi(\supp(V))\to\m F_p$ according to the commutative diagram:
\beas
\begin{CD}
\m f_n@>{\chi}>>\mathbb E_n\\
@V{V}VV @V{\mathcal V}VV\\
\m F_p @= \m F_p\\
\end{CD}
\eeas 
In this section we do not distinguish $V$ and its representation $\mathcal V$. This allows us operate with geometric images and their characteristics for rules and states. 

{\sl Diameter} $\delta(V)$ of a vector $V\in\m M[F_n,\m F_p]$ is the 
length of an edge of the minimal cube such that its edges are parallel to coordinate axes in $\mathbb E_n$ and it covers the set $\chi(\supp(V))$. 

{\sl Weight} $||V||$ of the vector is $|\supp(V)|$.\footnote{Another and more accurate version: maximal coordinate-wise size of the set $\chi(\supp(V))$.}

Elements $V\in\m M[\m f_n,\m F_p]$ and $f\in\m f_n$ given, let $V^{\la f\ra}$ denote a vector s.t. 
\beas
V^{\la f\ra}(g)=V(g-f), g\in\m f_n.
\eeas 
We see it as vector $V$ {\sl shifted} by $f$.\footnote{Warning: $V^{\la f\ra}$ is not a power of $V$ with exponent $\la f\ra$!}
From the definition directly we have
\begin{lem}\label{free-lem1}
$\left[V^{\la f)\ra}\right]^{\la g\ra}=V^{\la f+g\ra}$.
\end{lem}
For $V\in\m M[\m g,\m F_p]$ {\sl sparsity} of the vector $V$ is the number $\sigma(V)$ such that $\sigma(V)=\frac l2+1$ where $l$ is the length of an edge of the maximal (respectively inclusion) cube obeying the condition that if we place its center into any $x\in\chi(\supp(V))$ then no other element of $\chi(\supp(V))$ appears in the cube. 
 
\begin{lem}\label{bak-spar-size}
For any rule $R$, any state $V$, and non-negative integers $m,l$ it holds: \\
(i) $\sigma(\bak_p^m(R))=p^m\sigma(R)$; \\
(ii) $\delta(\bak_p^m(R))=p^m(\delta(R)-1)+1$;\\
(iii) $\delta(R^l*V)\le \delta(V)+l(\delta(R)-1)$.\\
In particular $ \delta(\bak_p^m(R))\le p^m\delta(R)$ and $\delta(R^l*V)\le l\delta(R)+\delta(V)$.  
\end{lem}
\pr This is because for a free group the baker transformation $\bak_p$ just shifts any element $(x_1,\dots,x_n)\in\chi(\supp(R))$ into a position $p(x_1,\dots,x_n)=(px_1,\dots,px_n)$. Therefore the distance in $\mathbb E_n$ between different elements $v,w$ of $\supp(\bak_p(R))$ increases by the factor $p$ comparing to the  distance between the preimages of the elements $v,w$. From here (i) and (ii) follow since the distances between the projections of  the elements from $\chi(\supp(R))$ on all axes increase by the same factor. 

On the basis of th edefinition of an application of a rule $R$ to a state $V$ and taking into account the finiteness of both $R,V$ it is not difficult to see that $\delta(R*V)\le \delta(V)+\delta(R)-1$. Indeed, 
\beas
R*V|^f=\sum_gV_gR^{-}_{f-g}=\sum_gV_gR_{g-f}
\eeas 
and when the maximal component $\lambda$ of $\chi(g-f)$ exceeds $\delta(Q)$ we get $R*V|^f=0$. Thus we can consider only cases when $\lambda\le\delta(V)+\delta(R)-1$. By a simple induction one can check that $\delta(R^l*V)\le \delta(V)+l(\delta(R)-1)$.
\bx

Thus, we can say that any application of $\bak_p$ {\sl inflates}  vectors.

We call vectors $V_i,i\in M,$ {\sl separated} from each other if the convex hulls of the sets $\{\chi(v)|v\in\supp(V_i)\},i\in M,$ have no common elements with each other in $\mathbb E_n$.

Below the following constructions with vectors from $\m M[\m f_n,\m F_p]$ are used. We say that a vector $Y$ is obtained by {\sl $p$-inflation} of order $m$ from a vector $X$ if $Y(p^mg)=X(g)$ and $\supp(Y)=\{p^mg|g\in\supp(X)\}$. 

Another operation is defined as follows. A vector $X$ and a vector $Y$ given, assume that for any different $y,y'\in\supp(Y)$ vectors $X^{\la y\ra},X^{\la y'\ra}$ are separated from each other. 
Then we define 
\beas
X\uparrow Y=\sum_{y\in\supp(Y)}Y(y)X^{\la y\ra}.
\eeas 
and call the vector $X\uparrow Y$ {\sl iteration} of the vector $X$ by the vector $Y$. It is clear that the vector $X\uparrow Y$ {\bf consists of $||Y||$ copies of the vector $X$ separated from each other}.   

From the definitions and lemma~\ref{bak-spar-size} it follows:
\begin{cor}\label{p-infl}
A free commutative group $\m f_n$, a modulus $p$, and a vector $X\in\m M[\m f_n,p]$ given, the vector $\bak_p^m(X)$ is a $p$-inflation of the vector $X$ of the order $m$. 
\end{cor}

A basic fact used below can be formulated as follows:
\begin{lem}\label{free-lem2}
Let $R,V$ be a rule and an initial state of an automaton. 
Assume $k=m+j+l$ where $l,j,m\in\mathbb Z^+$ and 
$\delta(V^{[l]})\le\sigma(R^j)$. 
Then for the automaton's state $V^{[k]}$ at time $k$ it holds\footnote{Recall: we use $R^m$ instead of $\debt Rm$ and this relates to other powers of rule $R$.}
\bea\label{fl2}
V^{[k]}=R^m*\left(\sum_{r\in\supp R} R^j(r)\cdot (V^{[l]})^{\la r\ra}\right)=R^m*\left (V^{[l]}\uparrow R^{j}\right).
\eea 
\end{lem}
\pr
We have
\beas 
V^{[k]}=R^k*V=R^m*(R^{j}*(R^l(V)))=R^m*(R^{j}*V^{[l]}).
\eeas 
Now, we can write
\beas
R^{j}=\sum_{r\in\supp(R^j)}R^j(r)K[r].
\eeas
Therefore by lemma~\ref{oper-bxt} (iii) we infer that 
\beas
R^{j}*V^{[l]}=\sum_{r\in\supp(R^j)}R^j(r)(K[r]*V^{[l]})= \sum_{r\in\supp(R^j)}R^j(r)(V^{[l]})^{\la r\ra}. 
\eeas 
The condition  $\delta(V^{[l]})\le\sigma(R^j)$ means that the distances between the elements of $\chi(\supp(R^{j}))$ are not lesser the size of the state $V^{[l]}$. Hence for different $r$ the supports of $(V^{[l]})^{\la r\ra}$ are separated from each other. 
From here we arrive at 
\beas
V^{[k]}=R^m*\left(V^{[l]}\uparrow R^{j}\right).\bxm 
\eeas

\begin{theo}\label{trace}
A rule $R$ and an initial state $V$ given, let $t=l+kp^i,k,l,i\in\mathbb Z^+,$ and $p^i\sigma(R^{k})\ge \delta(V^{[l]})$. Then $V^{[t]}$ is an iteration of $V^{[l]}$ by $R^{kp^i}$.  
\end{theo}
\pr 
To apply the previous lemma we set $m=0$ and prove that $\sigma(R^{kp^i})\ge p^i\sigma(R^k)$. For that we first represent $R^{kp^i}$ as $\underbrace{R^{k}\BT\dots\BT R^{k}}_{p^i\text{ times}} =\bak_p^i(R^k)$. As we know from the results about $\bak_p$ and  corollary~\ref{p-infl}  $(x_1,\dots,x_n)\in\chi(\supp(R^k))\iff p^i(x_1,\dots,x_n)\in\chi(\supp(\bak_p^i(R^k)))$. Because of lemma~\ref{bak-spar-size}(i) this means that $\sigma(R^{kp^i})= p^i\sigma(R^k)$. \bx

A simple sufficient condition for a state at a moment of time to be an iteration by a $\BT$-power of a rule is given by the next
\begin{cor}
$p^m\sigma(R^k)\ge l(\delta(R)-1)+\delta(V)\implies V^{[l+kp^m]}=V^{[l]}\uparrow R^{kp^m}$. 
\end{cor}
\pr Indeed on the basis of lemma ~\ref{bak-spar-size} we can write $p^m\sigma(R^k)\ge l(\delta(R)-1)+\delta(V)\ge\delta(V^{[l]})$. Now  theorem~\ref{trace} allows to conclude $V^{[l+kp^m]}=V^{[l]}\uparrow R^{kp^m}$.\bx

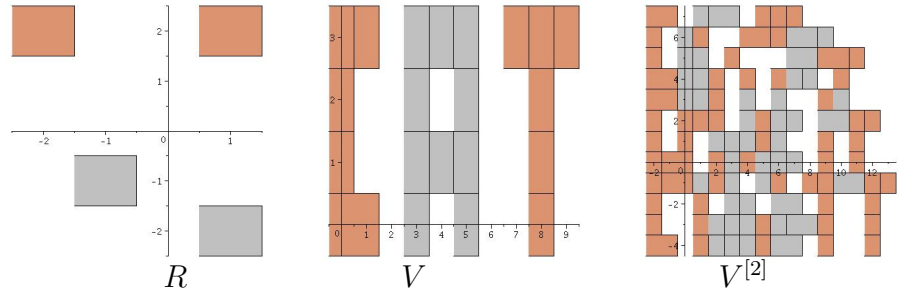
\begin{figure}[here]
 \begin{picture}(100,150)(0,0)
\put(-120,6){\includegraphics[width=3.5cm]{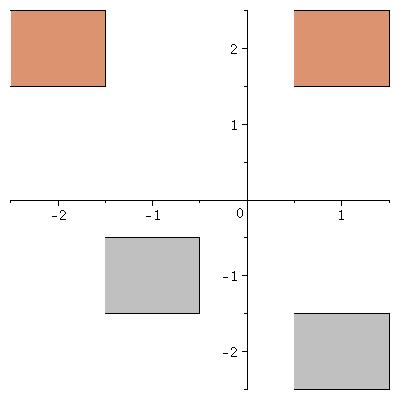}}
\put(0,6){\includegraphics[width=3.5cm]{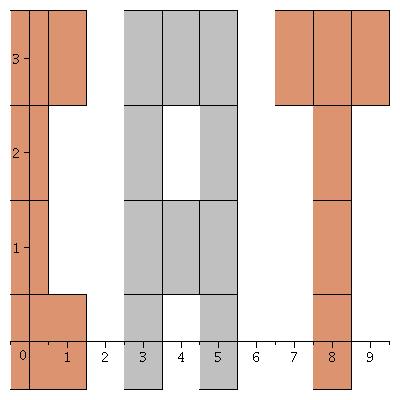}}
\put(120,6){\includegraphics[width=3.5cm]{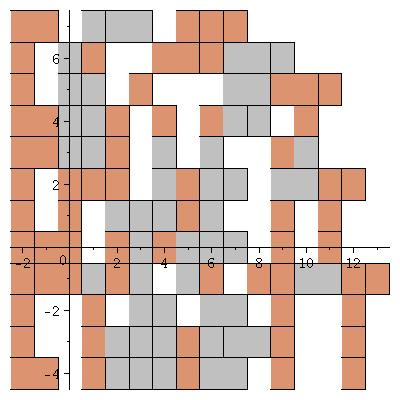}}
\put(-60,-3){$R$}\put(30,-3){$V$}\put(150,-3){$V^{[2]}$}
\end{picture}
\caption{Example~\ref{ex0}: Rule $R$, initial state $V$, and state $V^{[2]}$ at $t=2$. Gray color corresponds to value 2 of a cell state, brown - value 1 (modulus $p=3$).}\label{r4-cat}
\end{figure}

\begin{figure}[here]
 \begin{picture}(100,150)(0,0)
\put(-30,6){\includegraphics[width=3.5cm]{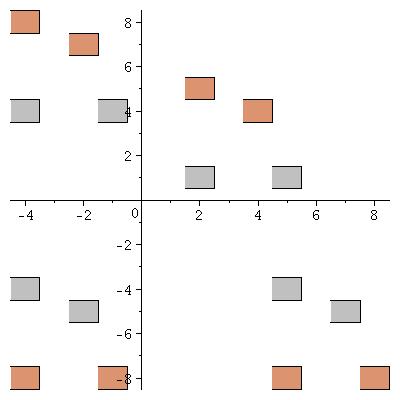}}
\put(90,6){\includegraphics[width=3.5cm]{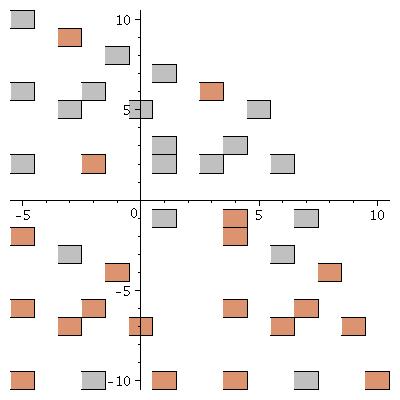}}
\put(10,-3){$(R^-)^{4}$}\put(120,-3){$(R^-)^{5}$}
\end{picture}
\caption{Example~\ref{ex0}: Powers $(R^-)^{4}$ and $(R^-)^{5}$.}\label{R4-R5}
\end{figure}

\begin{figure}[here]
 \begin{picture}(100,150)(0,0)
\put(-30,6){\includegraphics[width=3.5cm]{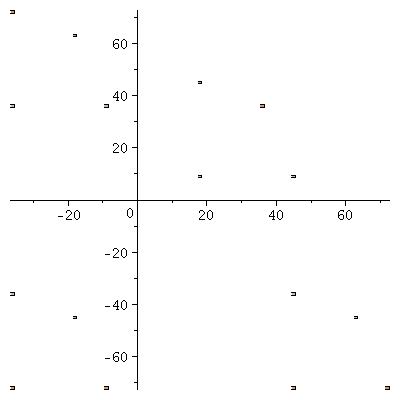}}
\put(90,6){\includegraphics[width=3.5cm]{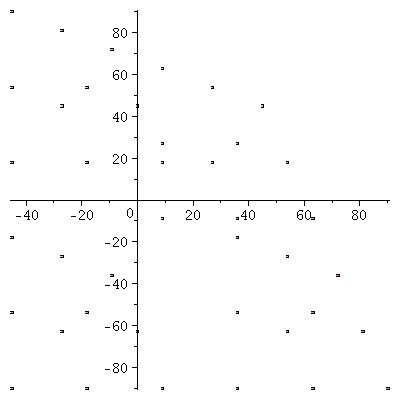}}
\put(10,-3){$(R-)^{36}$}\put(120,-3){$(R^-)^{45}$}
\end{picture}
\caption{Example~\ref{ex0}: Powers $(R^-)^{36}$ and $(R^-)^{45}$.}\label{R36-R45}
\end{figure}

\begin{figure}[here]
 \begin{picture}(100,150)(0,0)
\put(-60,6){\includegraphics[width=4cm]{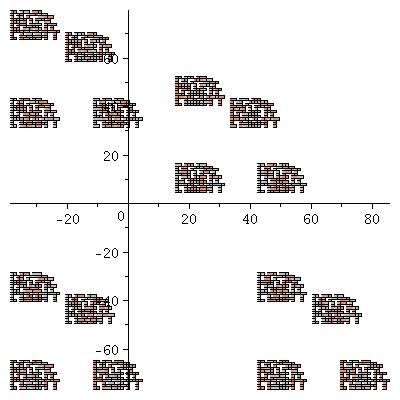}}
\put(90,6){\includegraphics[width=4cm]{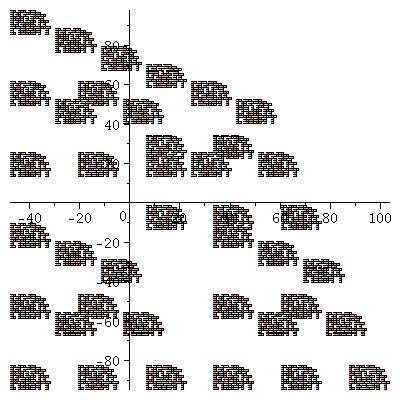}}
\put(-30,-3){$V^{[38]}$}\put(120,-3){$V^{[47]}$}
\end{picture}
\caption{Example~\ref{ex0}: states $V^{[38]}$ and $V^{[47]}$.}\label{v38-v47}
\end{figure}

\begin{ex}\label{ex0}{\rm
Let $p=3$. For the rule $R$ and the initial state $V$ shown on Fig.~\ref{r4-cat} we have $\delta(R)=5,\sigma(R)=2, \delta(V)=10$ whereas $\delta(V^{[2]})=16$. Also $\sigma(R^4)=2$ whereas $\sigma(R^5)=1$.
Let $l=i=2, k=4$. The powers $R^4, R^5$ represented in the symmetric form by $(R^-)^4, (R^-)^5$ on Fig~\ref{R4-R5}.

Then $p^i\sigma(R^4)=9\cdot2=18>16=\delta(V^{[2]})=16$. For this case $t=l+kp^i=38$ and theorem~\ref{trace} states that $V^{[38]}$ should be the iteration of $V^{[2]}$ (see Fig.~\ref{r4-cat}) by $R^{36}$ which is a $3$-inflation of $R^{4}$ of order 2. 
The rule $(R^-)^{36}$ (which is a symmetric to $R^{36}$) is shown on Fig.~\ref{R36-R45} and $V^{[38]}$ is shown on Fig.~\ref{v38-v47}  (left). 

The values $i=2,k=5,l=2$ define the time moment $t=l+kp^i=2+45=47=2+1200_3$ when the condition from theorem~\ref{trace} is not obeyed. Indeed $p^i\sigma(R^{5})=p^i=3^2<\delta(V[2])=16$. Therefore (see Fig.~\ref{v38-v47}, right part)  $V^{[47]}$ is not an iteration of $V^{[2]}$ by the rule $R^{45}$ despite $R^{45}$ is a $p$-inflation of $R^{5}$ of the order 2.  
}\bx
\end{ex}

One more 
\begin{cor}\label{inflation}
A rule $R\in\m M[\m f_n,\m F_p]$ and a number $m\in\mathbb Z^+$ given, for any initial state $V$ such that $p^m\sigma(R)\ge\delta(V)$ and for any integer $k\ge m$ 
\beas
V^{[p^k]}=V\uparrow R^{p^k} 
\eeas 
(i.e. the state $V^{[p^k]}$ is an iteration of $V$ by a $p$-inflation $R^{p^k}$ of the rule $R$ of the order $k$). 
\end{cor}
\pr Set $l=0$ in the previous corollary.\bx

The corollary states that if we observe at moments $t=p^k, k\ge m,$ the automaton on $\m f_n$ defined by a rule $R$ and an initial state $V$ such that $p^m\sigma(R)\ge\delta(V)$ is true, then in $\mathbb E_n$ we will see permanent movement of $||R||$ copies of the initial state $V$ away from each other as it is illustrated by the next two examples.
 \begin{figure}[here]
 \begin{picture}(100,150)(0,0)
\put(-30,6){\includegraphics[width=3.5cm]{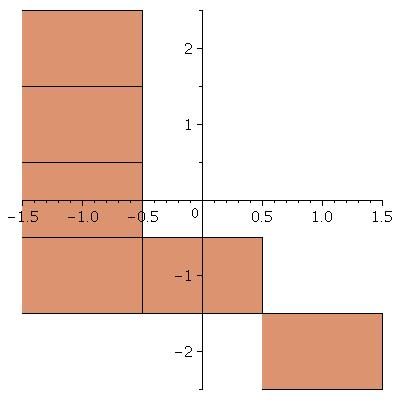}}
\put(90,6){\includegraphics[width=3.5cm]{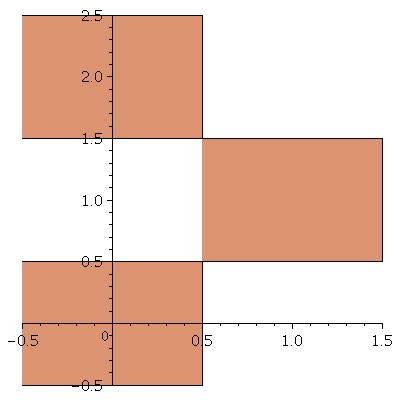}}
\put(20,-3){$R$}\put(120,-3){$V$}
\end{picture}
\caption{Rule $R$ and initial state $V$ (modulus $p=2$).}\label{r-v}
\end{figure}

 \begin{figure}[here]
 \begin{picture}(100,150)(0,0)
\put(-120,6){\includegraphics[width=3.5cm]{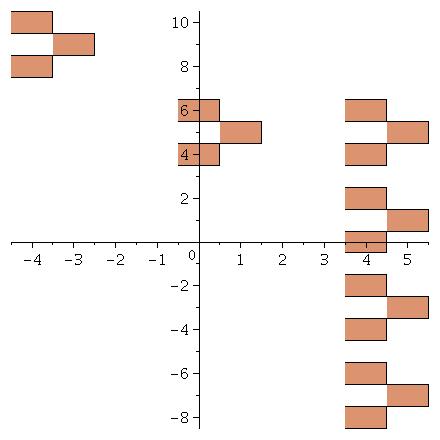}}
\put(20,6){\includegraphics[width=3.5cm]{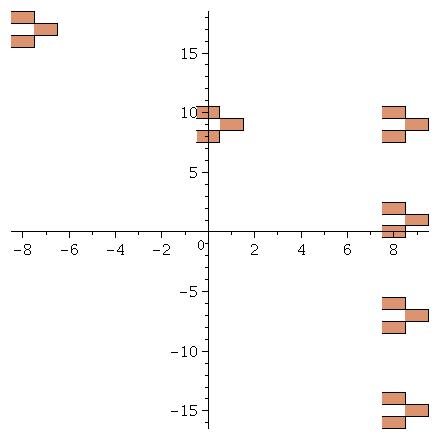}}
\put(160,6){\includegraphics[width=3.5cm]{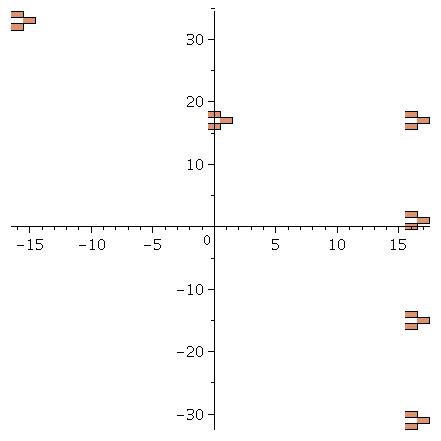}}
\put(-90,-3){$t=4$}\put(50,-3){$t=8$}\put(190,-3){$t=16$}
\end{picture}
\caption{$V^{[t]}$ for rule $R$ and initial state $V$ shown on Fig.~\ref{r-v}.}\label{2-3-4}
\end{figure}
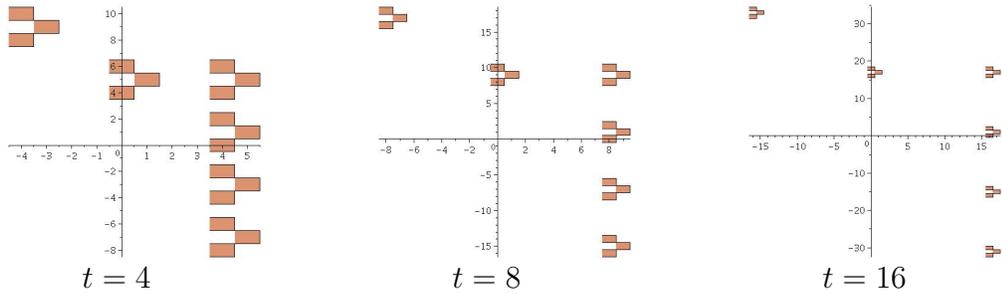
\begin{ex}\label{ex1}{\rm
Let $p=2$. For the rule $R$ and the intial state $V$ shown on Fig.~\ref{r-v} we have $\sigma(R)=1,\delta(V)=2$. Therefore for each $m\ge2$ it holds that $V^{[2^m]}$ is a 2-inflation of the order $m$ of the intitial state by the rule. Fig.~\ref{2-3-4} shows cases $m=2,3,4$.}\bx
\end{ex}

Corollary~\ref{inflation} suggests an upper estimate of the number $m$ (a threshold of inflation) starting from which we obtain inflation of the initial state for all moments of time $p^k,k\ge m$. However for some initial states $V$ obeying the condition of the corollary it could be that an inflation happens at moments $p^k$ for some $k<m$ as well. The next example demonstrate this.\footnote{More accurate estimate could be built in terms of coordinate-wise sizes for states and coordinate-wise sparsities for rules.} 

 \begin{figure}[here]
 \begin{picture}(100,150)(0,0)
\put(-30,6){\includegraphics[width=3.5cm]{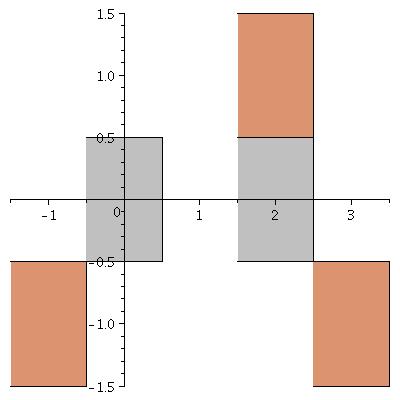}}
\put(90,6){\includegraphics[width=3.5cm]{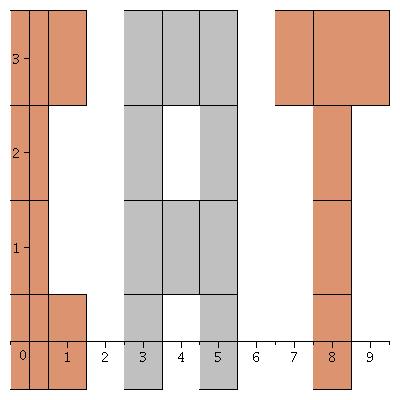}}
\put(20,-3){$R$}\put(120,-3){$V$}
\end{picture}
\caption{Rule $R$ and initial state $V$. Gray color corresponds to the value 2 of a cell state, brown - the value 1 (the modulus $p=3$).}\label{r-v-cat}
\end{figure}

 \begin{figure}[here]
 \begin{picture}(100,170)(0,0)
\put(-120,16){\includegraphics[width=3.5cm]{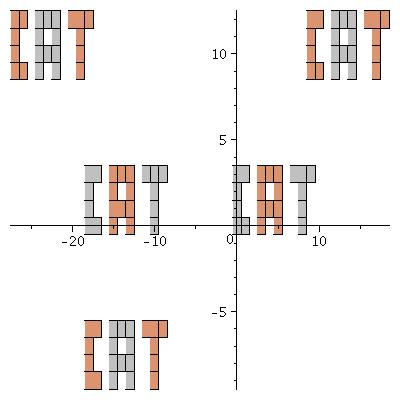}}
\put(-10,6){\includegraphics[width=4cm]{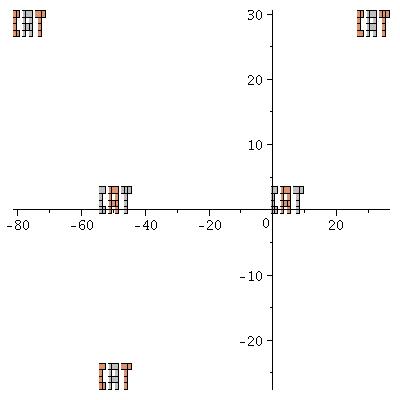}}
\put(110,6){\includegraphics[width=5.7cm]{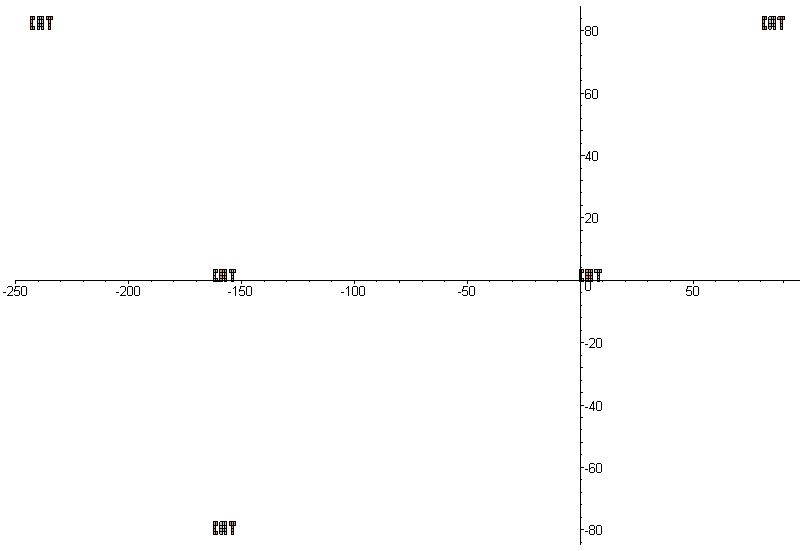}}
\put(-80,-3){$t=9$}\put(50,-3){$t=27$}\put(210,-3){$t=81$}
\end{picture}
\caption{$V^{[t]}$ for the rule $R$ and the initial state $V$ shown on Fig.~\ref{r-v-cat}.}\label{cat2-3-4}
\end{figure}
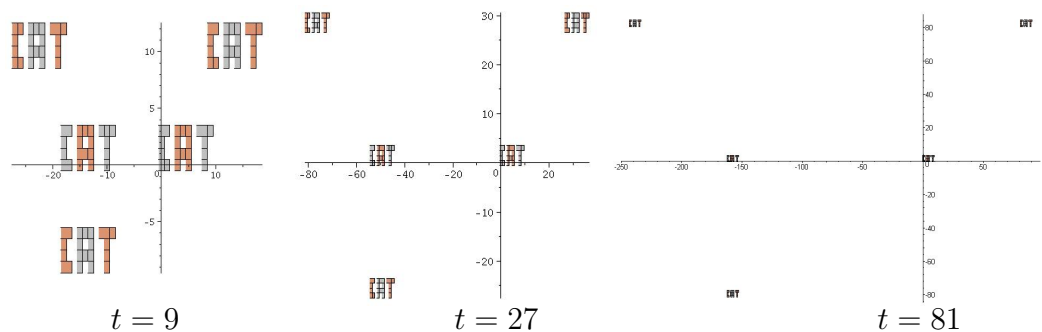

\begin{ex}\label{ex2}{\rm
Let $p=3$. For the rule $R$ and the intial state $V$ shown on Fig.~\ref{r-v-cat} we have $\sigma(R)=1,\delta(V)=10$. Therefore the estimate for $m$ given by corollary~\ref{inflation} is 3. However for each $m\ge2$ it holds that $V^{[3^m]}$ is a 3-inflation of the order $m$ of the intitial state by the rule. Fig.~\ref{cat2-3-4} shows cases $m=2,3,4$.}\bx
\end{ex}

In logarithmic scale for space and time the behavior described by corollary~\ref{inflation}
could be seen as a movement of $||R||$ copies of an initial state through the space away from each other. This reminds glider's flight (if we use terminology from the theory of  J.Conway's automaton {\sl Life}). Two essential differences are:
\begin{itemize} 
\item the movement of each copy (``glider'') is not an autonomous;
\item the behavior is a typical does not matter what are non-zero rule and initial state. 
\end{itemize}

\begin{theo}\label{iter-struct}
A rule $R$ and an initial state $V$ given, let $l_j=\sum_{i=1}^jp^{m_i}, 0<j\le k\in\mathbb Z,$  where $\sigma(R)p^{m_1}\ge\delta(V)$ and for each $j\in\{1,2,\dots,k-1\}$
it hold that $p^{m_{j+1}}>l_j\frac{\delta(R)-1}{\sigma(R)}$. Then 
\beas
V^{[l_k]}=(\dots(V\uparrow \bak_p^{m_1}(R))\uparrow\dots)\uparrow
\bak_p^{m_k}(R).
\eeas 
One particular choice for $m_j, j=\ovl{2,k},$ could be $m_j=m_1+(j-1)\mu$  where $p^{\mu}\ge\frac{\delta(R)-1}{\sigma(R)}+1.$
\end{theo}
\pr 
When $k=1$ this statement follows from theorem~\ref{trace}. Note that 
\beas 
\delta(V^{[l_j]})\le {l_j}(\delta(R)-1)+\delta(V)\le {l_j}(\delta(R)-1)+p^{m_1}\sigma(R),
\eeas 
lemma~\ref{bak-spar-size}(iii). Hence if 
\beas 
p^{m_{j+1}}\sigma(R)
\ge {l_j}(\delta(R)-1)+p^{m_1}\sigma(R)
\eeas  
then $V^{[l_{j+1}]}=V^{[l_j]}\uparrow R^{p^{m_{j+1}}}$. We can rewrite
\beas
p^{m_{j+1}}\sigma(R)\ge {l_j}(\delta(R)-1)+p^{m_1}\sigma(R)\quad\text{as}\quad 
p^{m_{j+1}}\ge {l_j}\frac{\delta(R)-1}{\sigma(R)}+p^{m_1}.
\eeas 
The latter inequality can be satisfied by the condition
\beas 
\frac{p^{m_{j+1}}}{l_j}>\frac{\delta(R)-1}{\sigma(R)}\quad\text{because}\quad \frac{p^{m_1}}{{l_j}}<1\quad\text{if}\quad j>1.
\eeas 
Thus the ``general part'' of the statement is proved. 

Assume $m_{j+1}=m_j+\mu,j=\in\mathbb Z^+$ and thereby $m_j=m_1+(j-1)\mu$. From here 
\beas
l_j=\sum_{i=1}^jp^{m_i}=p^{m_1}\sum_{i=0}^{j-1}(p^{\mu})^i=p^{m_1} \frac{p^{j\mu}-1}{p^{\mu}-1}
\eeas
Hence
\beas 
\frac{p^{m_{j+1}}}{l_j}=\frac{p^{m_{j+1}}(p^{\mu}-1)}{p^{m_1}(p^{j\mu}-1)} = 
\frac{p^{m_1}p^{j\mu}(p^{\mu}-1)}{p^{m_1}(p^{j\mu}-1)}>p^{\mu}-1
\eeas 
because $j>1,\mu\ge1,p\ge2$. This means that we can choose $\mu$ satisfying the condition $p^{\mu}\ge\frac{\delta(R)-1}{\sigma(R)}+1$. 
\bx\\

According to this theorem the arising iterational structures approximate in a sense self-similar structures at appropriate time moments. The following examples shows the structures of level of iteration 2, 3, and 4 at the first time moments of their appearance. 
\begin{figure}[here]
 \begin{picture}(180,180)(0,0)
\put(-120,16){\includegraphics[width=3.5cm]{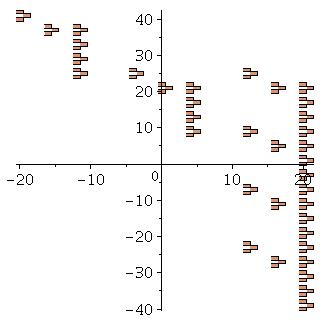}}
\put(-10,6){\includegraphics[width=4cm]{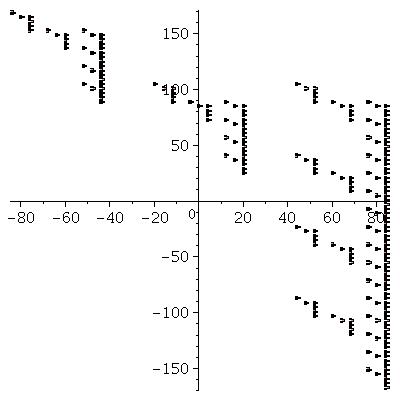}}
\put(110,6){\includegraphics[width=5.7cm]{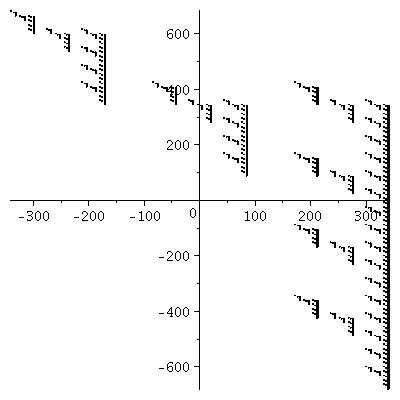}}
\put(-80,-3){$t=20$}\put(40,-3){$t=84$}\put(180,-3){$t=340$}
\end{picture}
\caption{Iterational structures for $p=2$ and $R,V$ shown on Fig.~\ref{r-v}.}\label{iter-str}
\end{figure}

\begin{ex}\label{iter1}{\rm
Fig.~\ref{iter-str} shows the iterational structures for the data ($p,R,V$) from example~\ref{ex1} at time moments $t=20,84,340$. The $p$-adic notations of these numbers are $10100,1010100,101010100$ respectively. (That is for instance $340=2^2+2^4+2^6+2^8$.) We have 
\beas 
V^{[340]}=\left[\left[\left[V\uparrow\bak_2^2(R)\right]\uparrow\bak_2^4(R)\right]
\uparrow\bak_2^6(R)\right]\uparrow\bak_2^8(R) 
\eeas
and each iteration $\uparrow$ from left to right in this formula iterates the previous structure by the rule $R$ appropriately inflated. This can lead (in a limit) to a certain kind self-similarity.

On the other hand, due to $\delta(R)=5,\sigma(R)=1$  theorem~\ref{iter-struct} warranties iteration structures similar to the presented on Fig.~\ref{iter-str} only when the difference $m_{i+1}-m_i$ is equal to 3 at least. Therefore the numbers $t=20,84,340$ should (according to the theorem) look like $t=36,292,2340$. 

This example shows that the characteristic $\delta(R)$ could be replaced with more accurate one. For example, as we noticed above, the further step could be a replacement of the defined sizes of rules and states  with collections of coordinate-wise sizes. In this case more accurate consideration shows that in fact an effective size of $R$ in this example is equal to 3. From here $2^{\mu}$ must be not lesser than 3 resulting in $\mu=2$. And the series of $m_i$ calculated on the basis $m_1=2$ and $\mu=2$ consists of exactly the same time moments $20,84,340,\dots$ as on the Fig.~\ref{iter-str}. 
}\bx
\end{ex}

\subsection{Cartesian products of commutative groups}\  \\

Here the decomposition effects are combinations of the effects described above for periodical groups and free groups because we consider only finitely generated commutative groups. 

One peculiarity that worthes to be noticed could be formulated as follows: the observable size of the periodical component $G$ for product $G\times F$ with a free group $F$ can depend on the size of the step of observations in time. In degenerated case when $G$ is a $p$-group with the maximal order of its elements $p^s$ and the length of the step is $p^k,k\ge s$, the component $G$ is unobservable. 
With more details we consider a similar case below for direct products of non-commutative $2$-groups of a certain kind with commutative groups.

\section{ACA on non-commutative groups for the case $p=2$}

\subsection{Representation of $\BT$ by operations on sets in case $p=2$}\ \\

Now we consider another representation of the convolution $\BT$.
A modulus $p$ and vectors $V,W\in\m M[\m g,\m F_p]$ given, let be 
\bea\label{circledstar}
V\circledast W=\left\{f\ \text{\bf :}\ \  
\sum_{\substack{v'\in\supp(V),w'\in\supp(W)\\
v'w'=f}}V(v')W(w')\ \neq0\mm p\right\}.
\eea
Note that the result of the operation $\circledast$ is not a vector but a subset of $\m g$.  
\begin{lem}\label{l4} 
 $\supp(V\BT W)=V\circledast W$. 
\end{lem}
\pr By the definition of $\BT$ and $\supp$ we have 
\beas 
f\in\supp(V\BT W)\iff\sum_{v\in\m g}V(v)W(\ovl vf)\neq0\mm p.
\eeas 
Let $w=\ovl vf$. 
\beas
f\in\supp(V\BT W)\iff\  
 \sum_{\substack{v',w'\in\m g\\f=v'w'}}V(v')W(w')\neq0\mm p.
\eeas
Taking into account that $f=vw$ and $X(u)\neq0\iff u\in\supp(X)$
we can rewrite the previous equivalence as following
\beas 
f\in\supp(V\BT W)\iff\  
\sum_{\substack{v'\in\supp(V),w'\in\supp(W)\\v'w'=f}}V(v')W(w')\neq0\mm p. \bxm
\eeas

\begin{cor}
The operation $\circledast$ is associative. In case when $A\BT B=B\BT A$ we have $A\cst B=B\cst A$.
\end{cor}
For the binary case ($p=2$) vectors actually are characteristic functions of sets and therefore the operation $\circledast$ can be seen as an operation on subsets of an index group and the definition~(\ref{circledstar}) could be rewritten as follows:
\beas
M\circledast L=\{f\ \text{\bf :}\ \left|\{(v',w')|v' \in M, w' \in L, v'w'=f\}|=1\mm 2\right\},\ \ M,L\subseteq\m g. 
\eeas
Further, in the case $p=2$ we can consider $\circledast$ as an operation on subsets of an index group. 
\begin{cor}\label{bxt}
If $p=2$ then  $\supp(A\BT B)=\supp (A)\circledast \supp (B)$.
\end{cor}
\pr
This is a direct consequence of lemma~\ref{l4} and  definition~(\ref{circledstar}).
\bx

One more useful tool for the case $p=2$ is an operation $\flat:\mathcal P(\m g)\to\mathcal P(\m g)$ on sets of elements of an index-group $\m g$ which accompanies the baker transformation. Namely for $G\subseteq\m g$ we define
\beas
\flat(G)=\{g\ \text{{\bf:}}\ g\in \m g\ \wedge\ 
|\{q|q\in Q\ \wedge \ q^2=g\}=1 | \mm 2 \}.
\eeas
As usual we accept that $\flat^0(G)=G$ and $\flat^{n+1}(G)=\flat(\flat^n(G))$. 

Let $Q_i,i=\ovl{1,n},$ be sets. We define 
\beas
\underset{i=1}{\overset{n}{\uplus}}Q_i=\{q\text{\bf:}\ \  q\in\cup_{i=1}^nQ_i\ \wedge\ |\{i|q\in Q_i\}|=1\mm 2\}.
\eeas 
For sets $Q,S$ we have $Q\uplus S=(Q \cup S)\setminus(Q\cap S)$.

In this section for a subset $Q$ of a group $\m g$ by $\ovl Q$ we denote the set $\{\ovl g|g\in Q\}$ where $\ovl g$ is the inverse element for $g$ in $\m g$.

We call a subset $S$ of an index group $\m g$ {\sl abelian set} if all elements of $S$ commute with each other in $\m g$.

\begin{cor}\label{flat}
$p=2$ given,\\
(i) for any $V\in\m M[\m g,\m F_2]$ such that $\supp(V)$ is an abelian set in $\m g$, it holds:
$$\sup(V\BT V)=\supp(V)\cst\supp(V)=\flat(\supp (V));$$
(ii) for any abelian sets $Q,S$
\beas 
\flat(Q\uplus S)=\flat(Q)\uplus\flat(S),\\
\flat(Q\cst S)=\flat(Q)\cst\flat(S),\\
\flat(\ovl Q)=\ovl{\flat(Q)}.
\eeas
\end{cor}
\pr
(i) Let $S=\supp(V)$. We need to show that 
\beas
S\cst S=_{\text{df}}\{ss'\text{\bf:}\ \  |\{s''s'''\text{\bf:}\ \ s'',s'''\in S,\  ss'=s''s'''\}|=1\mm 2\}=\\
\{s^2\text{\bf:}\ \ |\{(s')^2\text{\bf:}\ \ s'\in S,\ s^2=(s')^2\}|=1\mm 2\}.
\eeas 
This follows from the fact that each pair $(q,q')$ where $q\neq q', \ q,q'\in S$ should be counted twice and therefore only cases to take into account are those where $s=s', s''=s'''$.\\
(ii) This is because $s\in S\ \wedge\ q\in Q\implies sq=qs\ \wedge\ s^2q^2= (sq)^2\ \wedge\ (\ovl q)^2=\ovl{q^2}$. 
\bx

\subsection{ Groups $\mathrm{DHC}(\mbf S,\theta)$ of order $2^{k+1}$}\ \\

Group $\mathrm{DHC}(\mbf S,\theta)$ is built from a finite abelian group $\mbf S$ and its element $\theta$ of the order not exceeding 2 by extension with an element $r$ and defining relations: $r^2=\theta$ and $rs=\ovl sr$ for all $s\in\mbf S$. 

When $\theta=\m 1$ a group $\mathrm{DHC}(\mbf S,\theta)$ is actually a generalized dihedral group $\mathrm{Dih}(\mbf S)$ (see \cite{Dih2}, p. 210). And when $\mbf S$ is a cyclic group of order $q$ then $\mathrm{DHC}(\mbf S,\theta)$ is a dihedral group $\m D_q$ with representation $\la\{s,r\}|s^q=1,\  r^2=1,\ sr=rs^{-1}\ra$.

In case $\theta\neq1$ a group $\mathrm{DHC}(\mbf S,\theta)$ is a generalized dicyclic group $\mathrm{Dic}(\mbf S,\theta)$ (see \cite{Dick}, p. 170, \cite{Scott}, p. 252). 

If $\mbf S$ is a cyclic group of an even order, its element $\theta$ of the order 2 is defined uniquely and we arrive at a dicyclic group  $\mathrm{Dic}(\mbf S)$. Examples are: the quaternion group ($\mbf S$ is a cyclic group of order 4) and the generalized quaternion group ($\mbf S$ is cyclic group of order 8). 

Some sources (see \cite{Rot}, p. 96) define as a generalized quaternion group $\m Q_n$ any generalized dicyclic group $\mathrm{Dic}(\mbf S)$ when $\mbf S$ is a cyclic group of order $2^n, n\ge 3$. 

Due to the relations $rs=\ovl sr$ for any $s\in\mbf S$ the center $\m Z$ of $\mathrm{Dic}(\mbf S)$ includes all elements of $\mbf S$ of the order not exceeding 2. In particular, $\theta\in\m Z$.

\begin{theo}\label{gendicyc}
Let $\mbf S$ be a commutative group of an order $2^k,k\in\mathbb Z^+,$ and $2^{\kappa}$ be the maximal order of elements of $\mbf S$. For any vector $V\in\m M[\mathrm{DHC}(\mbf S,\theta),\m F_2]$ it is true that $V^{\BT 2^{\kappa}}$ either is the unit of $M[\m D_{2^{k}},\m F_2]$ or has all components equal to zero.
\end{theo}
\begin{lem}\label{l5gc}
$(\forall V\in\m M[\mathrm{DHC}(\mbf S,\theta),\m F_2])(\exists!\ C,D\subseteq \mbf S)[\supp(V)=rC\cup D].$ 
\end{lem}

\pr From the relations $sr=r\ovl s, s\in\mbf S,$ and $r^2=\theta\in\mbf S$, it follows that any element of $\mathrm{DHC}(\mbf S,\theta)$ either belongs to $\mbf S$ or has representation $rs, s\in \mbf S$. \bx

\begin{lem}\label{l6gc}
Let $V$ be any fixed vector and $\supp(V)=rC_0\cup D_0$ whereas 
$\supp(V^{\BT\, 2^n})=rC_n\cup D_n$. The following recursion holds:
\beas
\begin{cases}
C_n=C_{n-1}\cst(D_{n-1}\uplus\ovl{D_{n-1}})\\
D_{n}=\flat(D_{n-1})\uplus \theta(C_{n-1}\cst\ovl{C_{n-1}}). 
\end{cases}
\eeas
\end{lem}
\pr
Induction on $n$. For $n\ge 1$ we have 
\beas
\supp (V^{\BT\,2^n})=\supp(V^{\BT\,2^{n-1}}\BT V^{\BT\,2^{n-1}})=
\supp(V^{\BT\,2^{n-1}})\cst\supp(V^{\BT\,2^{n-1}})=_{[\text{lemma~\ref{l4}}]}\\
(rC_{n-1}\cup D_{n-1})\cst(rC_{n-1}\cup D_{n-1})= \\
([rC_{n-1}\cst D_{n-1}]\uplus [D_{n-1}\cst rC_{n-1}])\cup ([rC_{n-1}\cst rC_{n-1}]\uplus [D_{n-1}\cst D_{n-1}]) =\\
rC_n\cup D_n  
\eeas 
where $C_n=C_{n-1}\cst(D_{n-1}\uplus\ovl{D_{n-1}})$ because $D_{n-1}\cst rC_{n-1}=r\ovl {D_{n-1}}\cst C_{n-1}=rC_{n-1}\cst \ovl{D_{n-1}}$: $\ovl{D_{n-1}}$ (as well as $D_{n-1}$)  comutes with $C_{n-1}$ since both are subsets of $\mbf S$. 

For $D_n$ we (due to $C_{n-1}r=r\ovl C_{n-1}$ and $r^2=\theta$) have 
$D_n=\theta C_{n-1}\cst\ovl{C_{n-1}}\uplus \flat(D_{n-1})$ 
because in abelian subgroup $\mbf S$ for any set $M$ we have $M\cst M=\flat(M)$ (corollary~\ref{flat}). 
\bx

\begin{lem}\label{l7gc}
Solution to recursion from lemma~\ref{l6gc} for $C_n,D_n$ is:
\beas
\begin{cases}
C_n=C_0\cst\,\underset{i=0}{\overset{n-1}{\circledast}}(\flat^{i}(D_0)\uplus\ovl{\flat^i(D_0)}), \\
D_n=\flat^{n}(D_0)\uplus\underset{i=0}{\overset{n-1}{\uplus}}\theta^{2^i} 
\left\{\flat^i(C_0)\cst\ovl{\flat^i(C_0)}\cst\underset{j=i+1}{\overset{n-1}{\circledast}}(\flat^{j}(D_0)\uplus\ovl{\flat^j(D_0)})\right\},
\end{cases}
\eeas
where it is assumed that $\underset{i=a}{\overset{b}{\circledast}}M_i=\{\m 1\}$ in case $a>b$.
\end{lem}

\pr
A parallel induction on $n$. For $n=1$ the statement of the lemma is obvious taking into account the agreements about $\flat^0$ and $\underset{a}{\overset{b}{\circledast}}$ with $a>b$.

Now we note that in the expression for $D_{n-1}$ the set $$M_{n-1}=\underset{i=0}{\overset{n-2}{\uplus}}\theta^{2^i} 
\left\{\flat^i(C_0)\cst\ovl{\flat^i(C_0)}\cst\underset{j=i+1}{\overset{n- 2}{\circledast}}(\flat^{j}(D_0)\uplus\ovl{\flat^j(D_0)})\right\}$$ is symmetric, that is 
$g\in M_{n-1}\implies \ovl g\in M_{n-1}$. In other words $M_{n-1}= \ovl{M_{n-1}}$. From here $M_{n-1}\uplus\ovl{M_{n-1}}= \emptyset$. Hence 
\beas
D_{n-1}\uplus\ovl{D_{n-1}}=\flat^{n-1}(D_0)\uplus\ovl{\flat^{n-1}(D_0)}.
\eeas 
Using the recursion for $C_n$ and the inductive hypothesis we arrive at
\beas
C_n=C_{n-1}\cst(D_{n-1}\uplus\ovl{D_{n-1}})=\\
C_{n-1}\cst(\flat^{n-1}(D_0)\uplus\ovl{\flat^{n-1}(D_0)})=\\
C_n=\left[C_0\cst\,\underset{i=0}{\overset{n-2}{\circledast}}(\flat^{i}(D_0)\uplus\ovl{\flat^i(D_0)})\right]\cst(\flat^{n-1}(D_0)\uplus\ovl{\flat^{n-1}(D_0)}).
\eeas 
For $D_n$ we replace $D_n$ in the recursion for $C,D$ with the expression according to inductive hypothesis
\beas
D_n=\flat(D_{n-1})\uplus\,\theta(C_{n-1}\cst\ovl{C_{n-1}})=\\
\flat\left(\flat^{n-1}(D_0)\uplus\underset{i=0}{\overset{n-2}{\uplus}} 
\theta^{2^i}\left\{\flat^i(C_0)\cst\ovl{\flat^i(C_0)}\cst\underset{j=i+1}{\overset{n-2}{\circledast}}(\flat^{j}(D_0) \uplus\ovl{\flat^j(D_0)})\right\}\right)\\ \uplus\,\theta(C_{n-1}\cst\ovl{C_{n-1}})=_{\text{lemma~\ref{flat}(ii)}}\\
\flat^{n}(D_0)\uplus\underset{i=0}{\overset{n-2}{\uplus}} 
\flat\left(\theta^{2^i}\left\{\flat^i(C_0) \cst\ovl{\flat^i(C_0)}\cst\underset{j=i+1}{\overset{n-2}{\circledast}}(\flat^{j}(D_0) \uplus\ovl{\flat^j(D_0)})\right\}\right)\\ \uplus\,\theta(C_{n-1}\cst\ovl{C_{n-1}})=_{\text{lemma~\ref{flat}(ii)}}\\
\flat^{n}(D_0)\uplus\underset{i=0}{\overset{n-2}{\uplus}}\theta^{2^{i+1}} 
\left\{\flat^{i+1}(C_0)\cst\ovl{\flat^{i+1}(C_0)}\cst\underset{j=i+1}{\overset{n-2}{\circledast}}(\flat^{j+1}(D_0)\uplus\ovl{\flat^{j+1}(D_0)})\right\}\\ \uplus\,\theta(C_{n-1}\cst\ovl{C_{n-1}})=\\
\flat^{n}(D_0)\uplus\underset{i=1}{\overset{n-1}{\uplus}}\theta^{2^{i}} 
\left\{\flat^{i}(C_0)\cst\ovl{\flat^{i}(C_0)}\cst\underset{j=i}{\overset{n-2}{\circledast}}(\flat^{j+1}(D_0)\uplus\ovl{\flat^{j+1}(D_0)})\right\} \uplus\theta(C_{n-1}\cst\ovl{C_{n-1}})=\\
\flat^{n}(D_0)\uplus\underset{i=1}{\overset{n-1}{\uplus}}\theta^{2^{i}} 
\left\{\flat^{i}(C_0)\cst\ovl{\flat^{i}(C_0)}\cst\underset{j=i+1}{\overset{n-1}{\circledast}}(\flat^{j}(D_0)\uplus\ovl{\flat^{j}(D_0)})\right\} \uplus\,\theta(C_{n-1}\cst\ovl{C_{n-1}}).
\eeas 
As we already know 
\beas
\theta\left(C_{n-1}\cst\ovl{C_{n-1}}\right)=\\
\theta\left(\left[C_0\cst\,\underset{i=0}{\overset{n-2}{\circledast}}(\flat^{i}(D_0)\uplus\ovl{\flat^i(D_0)})\right]\cst \ovl{\left[C_0\cst\,\underset{i=0}{\overset{n-2}{\circledast}}(\flat^{i}(D_0)\uplus\ovl{\flat^i(D_0)})\right]}\right)=_{\text{lemma~\ref{flat}}}\\
\theta\left(C_0\cst\ovl{C_0}\cst\,\underset{i=0}{\overset{n-2}{\circledast}}\left[(\flat^{i}(D_0)\uplus\ovl{\flat^i(D_0)})\cst\ovl{(\flat^{i}(D_0)\uplus\ovl{\flat^i(D_0)})}\right]\right)=\\ \theta\left(C_0\cst\ovl{C_0}\cst\,\underset{i=0}{\overset{n-2}{\circledast}}\flat(\flat^{i}(D_0)\uplus\ovl{\flat^i(D_0)})\right)=
\theta\left(C_0\cst\ovl{C_0}\cst\,\underset{i=0}{\overset{n-2}{\circledast}}(\flat^{i+1}(D_0)\uplus\ovl{\flat^{i+1}(D_0)})\right)=\\ \theta\left(C_0\cst\ovl{C_0}\cst\,\underset{i=1}{\overset{n-1}{\circledast}}(\flat^{i}(D_0)\uplus\ovl{\flat^{i}(D_0)})\right)=
\theta\left(\flat^0(C_0)\cst\flat^0(\ovl{C_0})\cst\,\underset{i=1}{\overset{n-1}{\circledast}}(\flat^{i}(D_0)\uplus\ovl{\flat^{i}(D_0)})\right). 
\eeas 
It is not difficult to see that the latter expression is the term of the sum
\beas
\underset{i=0}{\overset{n-1}{\uplus}}\theta^{2^{i}} 
\left\{\flat^{i}(C_0)\cst\ovl{\flat^{i}(C_0)}\cst\underset{j=i+1}{\overset{n-1}{\circledast}}(\flat^{j}(D_0)\uplus\ovl{\flat^{j}(D_0)})\right\}
\eeas 
corresponding to case $i=0$. This ends the proof of the solution of the recursion for $D_n$.
\bx

Let $\mbf S(j)$ be a subgroup of $\mbf S$ consisting of all elements from $\mbf S$ whose orders does not exceed $2^j$. 
\begin{lem}\label{l8gc}
It holds:\\
(i) $\mbf S(0)=\{\m1\},\quad\mbf S(1)\subseteq \m Z, 
\quad \mbf S(\kappa)=\mbf S$;\\
(ii) $\forall j[\,\mbf S(j)\le\mbf S(j+1)\,];$ \\
(iii) $\forall j\in[1,\kappa]\,\forall S\subseteq\mbf S(j)\,[\,\,\flat(S)\subseteq\mbf S(j-1)\,]$.  
\end{lem}

\pr Obvious. \bx

Since $\flat^{\kappa-1}(D_0)\subseteq\m Z$ and each element $z\in\m Z$ satisfies $z=\ovl z$, we get $\flat^{\kappa-1}(D_0)\uplus\ovl{\flat^{\kappa-1}(D_0)}=\emptyset$ and therefore $C_{\kappa}=\emptyset$ as well. 

By the same reason all terms in the sum 
\beas
\underset{i=0}{\overset{{\kappa}-1}{\uplus}} \theta^{2^i}
\left\{\flat^i(C_0)\cst\ovl{\flat^i(C_0)}\cst\underset{j=i+1}{\overset{{\kappa}-1}{\circledast}}(\flat^{j}(D_0)\uplus\ovl{\flat^j(D_0)})\right\}
\eeas 
are empty sets. This implies $D_{\kappa}=\flat^{\kappa}(D_0)\subseteq\{\m1\}$. Thus $\supp(V^{\BT2^{\kappa}})\subseteq\{\m 1\}$. This means that $V^{\BT2^{\kappa}}$ is either zero $\mbf 0$ of the semigroup or its unit $\mbf I$.
\bx\ \\

\section{Direct product of commutative group and groups $\mathrm{DHC}(\mbf S,\theta)$ of order $2^{k+1}$}

\begin{lem}\label{decart}
 Class  $\mathrm{DHC}(\mbf S,\theta)$ is closed under direct product. 
\end{lem}

\pr Let $G_i,i=1,2,$ be two groups of the class $\mathrm{DHC}(\mbf S,\theta)$ and for both values of $i$ the group $G_i$ is an extension of the abelian group $\mathbf{S_i}$ with its selected element $\theta_i$ whose order does not exceed 2. The extension is done with element $r_i$ obeying relations $r_i^2=\theta_i$ and  $r_is=\ovl sr_i$ for all $s\in\mbf S_i$. It is easy to see that $G_1\times G_2$ is a group of the class $\mathrm{DHC}(\mbf{S},\theta)$ with the specification $\mbf S=\mbf S_1\times\mbf S_2, \theta=(\theta_1,\theta_2)$, and for element $(r_1,r_2)$ it holds that $(r_1,r^2)^2=(r_1^2,r^2_2)=(\theta_1,\theta_2),\ (r_1,r_2)(s,s')=(r_1s,r_2s')=(\ovl sr_1,\ovl{s'}r_2)=\ovl{(s,s')}\,(r_1,r_2)$.
\bx

So we can restrict ourself with direct products $G\times A$ of one finite non-commutative group $G$ of sort $\mathrm{DHC}(\mbf S,\theta)$ and a commutative group $A$. 

For $V\in\m M[G\times A,\m F_2]$ let $\proj_G(V)$ denote $\{g\in G\text{ \bf : } \exists a\in A[(g,a)\in\supp(V)]\}$.

\begin{theo}\label{degeneration}
Let $A$ be a commutative group and $G$ be a 2-group of the class $\mathrm{DHC}(\mbf S,\theta)$ where the maximal order of elements of $\mbf S$ is $2^{\kappa}$.  For any rule $R\in\m M[G\times A,\m F_2]$ if $t=0\mm{2^{\kappa}}$ then $\proj_G(R^{\BT2^{\kappa}}) \subseteq\{\m1_G\}$ where $\m1_G$ is the unit of $G$.
\end{theo}

\pr
The proof is a modification of the proof given for theorem~\ref{gendicyc} because the statement from lemma~\ref{l5gc} holds here as well where group $\mbf S\times A$ plays the same role as $\mbf S$ before (we denote it by $\mathbb S$):
\begin{lem}\label{l5gc-decart}
$(\forall R\in\m M[G\times A,\m F_2])(\exists!\ C,D\subseteq \mathbb S)[\supp(R)=rC\cup D].$ 
\end{lem}

\pr 
This time we represent elements of $G\times A$ by pairs $(g,a),g\in G,a\in A$. 
Let $\m1_S,\m1_A,\m1_{\mathbb S}$ be units of groups $\mbf S,A,\mathbb S$ respectively and thereby $\m1_{\mathbb S}=(\m1_S,\m1_A)$. Clearly $\m1_{\mathbb S}=\m1_{G\times A}$.

For elements $r,\theta$ from the definition of a group of kind $\mathrm{DHC}(\mbf S,\theta)$ applied to $G$ we build elements $\mbf r,\Theta$ as $(r,\m1_A),(\theta,\m1_A)$ respectfully. We have of course: $s\mbf r=\mbf r\ovl s,s\in\mathbb S$, and $\mbf r^2=\Theta\in\mathbb S$. (Note that $\mbf r\notin\mathbb S$.) 
From here it follows that any element of $G\times A$ either belongs to $\mbf S\times A$, that is $\mathbb S$, or has representation $\mbf rs, s\in \mbf S\times A$, i.e. belongs to $\mbf r\mathbb S$. \bx

Thus sets $C_i,D_i$ from the proof of theorem~\ref{gendicyc} consist of pairs $(g,a)$ with component-wise multiplication as it is standard for direct products
of groups. The operations $\cst, \uplus,
\flat $ are directly applicable for sets of pairs that are subsets of group $\mathbb S$. Lemma~\ref{l4} and corollary~\ref{bxt} are applicable to subsets $S,Q$ of commutative group $\mathbb S$. 
For corollary~\ref{flat} we complete part (ii) with statement that for $Q\subseteq\mathbb S$ it holds $\flat(\mind Q)=\mind{\flat(Q)}$ where $\mind L$ denotes $\{(\ovl g,a)|(g,a)\in L\},\  L\subseteq\mathbb S$.
We need $\mind L$ because $(Q\times H)\mbf r=\mbf r(\ovl Q\times H)$  for $Q\subseteq\mbf S, H\subseteq A$. Thus $L\mbf r=\mbf r\mind L$.

The following lemma generalizes lemma~\ref{l6gc}.
\begin{lem}\label{l6gc-decart}
Let $V$ be any fixed vector and $\supp(V)=\mbf rC_0\cup D_0$ whereas 
$\supp(V^{\BT\, 2^n})=\mbf rC_n\cup D_n$. The following recursion holds:
\beas
\begin{cases}
C_n=C_{n-1}\cst(D_{n-1}\uplus\mind{D_{n-1}})\\
D_{n}=\flat(D_{n-1})\uplus \Theta(C_{n-1}\cst\mind{C_{n-1}}). 
\end{cases}
\eeas
\end{lem}
\pr
An induction on $n$. For $n\ge 1$ we have 
\beas
\supp (V^{\BT\,2^n})=\supp(V^{\BT\,2^{n-1}}\BT V^{\BT\,2^{n-1}})=
\supp(V^{\BT\,2^{n-1}})\cst\supp(V^{\BT\,2^{n-1}})=_{[\text{lemma~\ref{l4}}]}\\
(rC_{n-1}\cup D_{n-1})\cst(rC_{n-1}\cup D_{n-1})= \\
([rC_{n-1}\cst D_{n-1}]\uplus [D_{n-1}\cst rC_{n-1}])\cup ([rC_{n-1}\cst rC_{n-1}]\uplus [D_{n-1}\cst D_{n-1}]) =\\
rC_n\cup D_n  
\eeas 
where $C_n=C_{n-1}\cst(D_{n-1}\uplus\ovl{D_{n-1}})$ because $D_{n-1}\cst \mbf rC_{n-1}=r\mind {D_{n-1}}\cst C_{n-1}=\mbf r(\mind {D_{n-1}}\cst C_{n-1})=\mbf r(C_{n-1}\cst \ovl{D_{n-1}}) =\mbf rC_{n-1}\cst \ovl{D_{n-1}}$. We used the fact that $\ovl{D_{n-1}}$ (as well as $D_{n-1}$)  comutes with $C_{n-1}$ since both are subsets of $\mathbb S$. 

For $D_n$ we (due to $C_{n-1}\mbf r=\mbf r\mind {C_{n-1}}$ and $\mbf r^2=\Theta$) have 
$D_n=\Theta C_{n-1}\cst\mind{C_{n-1}}\uplus \flat(D_{n-1})$ 
because in abelian subgroup $\mathbb S$ for any set $M$ we have $M\cst M=\flat(M)$ (corollary~\ref{flat}). 
\bx

Further, lemma~\ref{l7gc} holds in form
\begin{lem}\label{l7gc-decart}
Solution to recursion from lemma~\ref{l6gc-decart} for $C_n,D_n$ is:
\beas
\begin{cases}
C_n=C_0\cst\,\underset{i=0}{\overset{n-1}{\circledast}}(\flat^{i}(D_0)\uplus\mind{\flat^i(D_0)}), \\
D_n=\flat^{n}(D_0)\uplus\underset{i=0}{\overset{n-1}{\uplus}}\Theta^{2^i} 
\left\{\flat^i(C_0)\cst\mind{\flat^i(C_0)}\cst\underset{j=i+1}{\overset{n-1}{\circledast}}(\flat^{j}(D_0)\uplus\mind{\flat^j(D_0)})\right\},
\end{cases}
\eeas
where it is assumed that $\underset{i=a}{\overset{b}{\circledast}}M_i=\{\m 1\}$ in case $a>b$.
\end{lem}
\pr The transformations from the proof of lemma~\ref{l7gc} could be repeated for $D_0,C_0\subseteq\mathbb S$ with replacement of $\ovl{\phantom{aa}}, r,\theta$ by $\mind{\phantom{aa}},\mbf r,\Theta$ respectively. \bx

The final part of the proof is based on lemma~\ref{l8gc} and the following fact to formulate which we denote $\{s|\exists a\in A[(s,a)\in Q]\}$ where $Q\subseteq\mathbb S$ by $\proj_SQ$.
\begin{lem}\label{lastlem}
If $\proj_S\left(\flat^m(D_0)\right)\subseteq\m Z$ then $\flat^m(D_0)\uplus\mind{\flat^m(D_0)}=\emptyset$.
\end{lem}
\pr First we notice that $s\in\m Z(\mbf S))\implies s^2=\m1_S$. Indeed, on one hand we have $rs=sr$ when $s\in\m Z$, on the other hand $\ovl sr=rs$. 

Since $\flat^m(D_0)$ and $\mind{\flat^m(D_0)}$ are sets and for each $(s,a)\in\flat^m(D_0)$, there exists $(\ovl s,a)\in \mind{\flat^m(D_0)}$. And yet $\ovl s=s$.  \bx

Thus $C_{\kappa}=\emptyset$ and all terms in sum 
\beas
\underset{i=0}{\overset{{\kappa}-1}{\uplus}} \Theta^{2^i}
\left\{\flat^i(C_0)\cst\mind{\flat^i(C_0)}\cst\underset{j=i+1}{\overset{{\kappa}-1}{\circledast}}(\flat^{j}(D_0)\uplus\mind{\flat^j(D_0)})\right\}
\eeas 
are empty sets. This implies $\proj_G(D_{\kappa})= \proj_S(D_{\kappa})=\flat^{\kappa}(\proj_S(D_0))\subseteq\{\m1_S\}$. Thus $\proj_G(\supp(R^{\BT2^{\kappa}}))\subseteq\{\m 1_G\}$, i.e.   $R^{\BT2^{\kappa}}$ is either zero $\mbf 0$ or $\proj_G(\supp(R^{\BT2^{\kappa}}))=\{\m1_G\}$. This means that the rule $R^{\BT2^{\kappa}}$ acts only on subgroup $A$.
\bx

\begin{cor}\label{vanishing}
Let $A$ be a commutative group and $G$ - a 2-group of the class $\mathrm{DHC}(\mbf S,\theta)$ where the maximal order of elements of $\mbf S$ is $2^{\kappa}$.  For any intial state $V$ and rule $R$ if $t=0\mm{2^{\kappa}}$ then $\proj_G(V^{[t]})$ either is equal to $\proj_G(V)$ or is empty.
\end{cor}

\begin{figure}[here]
 \begin{picture}(100,150)(0,0)
\put(-90,16){\includegraphics[width=3.5cm]{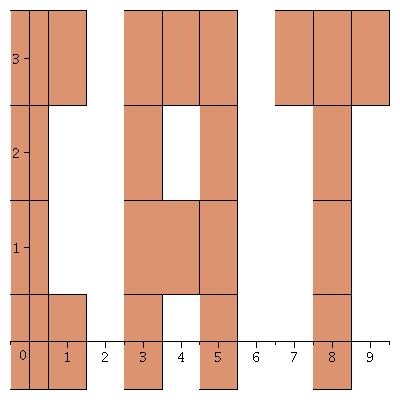}}
\put(30,6){\includegraphics[width=4.5cm]{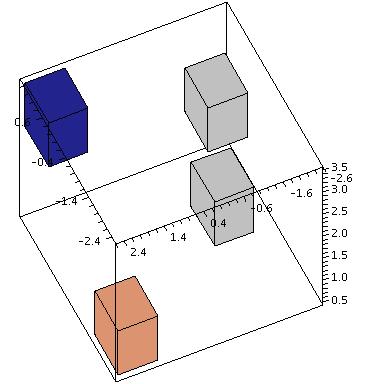}}
\put(-60,-5){$V$}\put(60,-5){$R$}
\end{picture}
\caption{Example~\ref{3-d-cat}: Initial state $V$ and rule $R$ (modulus $p=2$).}\label{quat1}
\end{figure}

\begin{figure}[here]
 \begin{picture}(100,150)(0,0)
\put(-90,-10){\includegraphics[width=4.5cm]{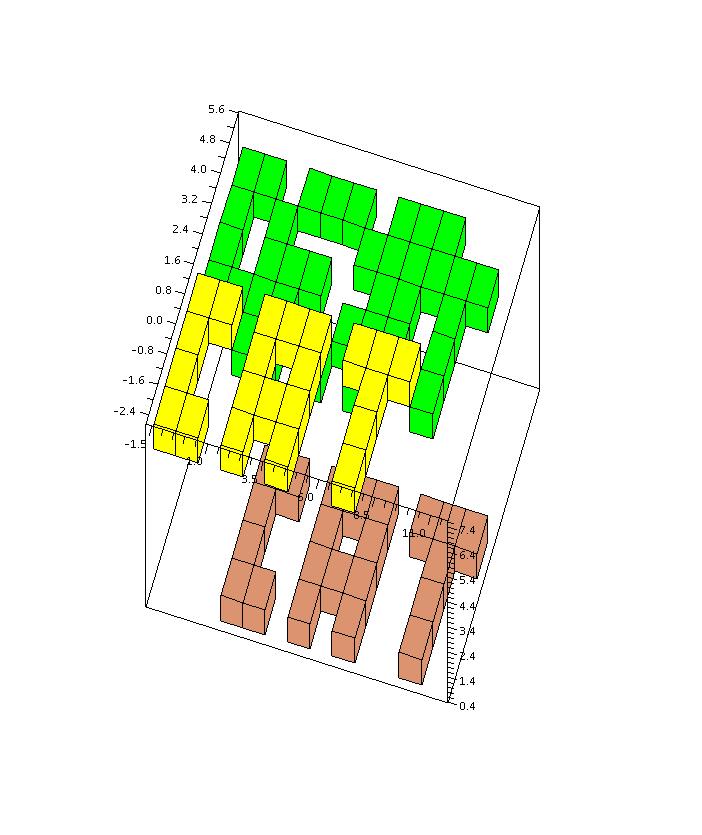}}
\put(90,16){\includegraphics[width=3.5cm]{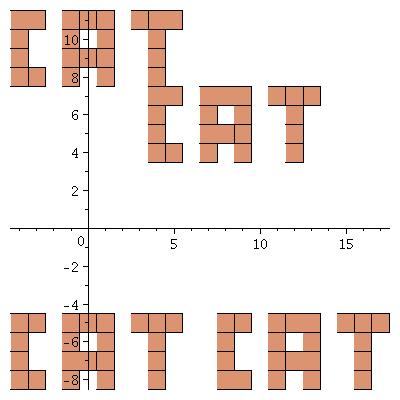}}
\put(-40,-5){$V^{[1]}$}\put(120,-5){$V^{[4]}$}
\end{picture}
\caption{Example~\ref{3-d-cat}: States on step 1 (left) and 4 (right).}\label{quat2}
\end{figure}

\begin{figure}[here]
 \begin{picture}(100,150)(0,0)
\put(-60,0){\includegraphics[width=4.5cm]{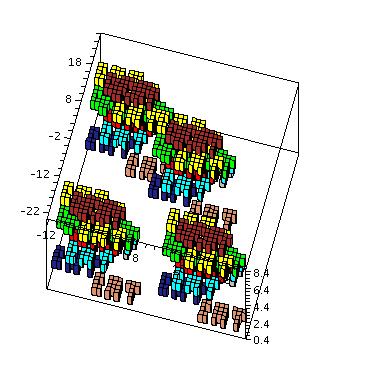}}
\put(90,16){\includegraphics[width=3.5cm]{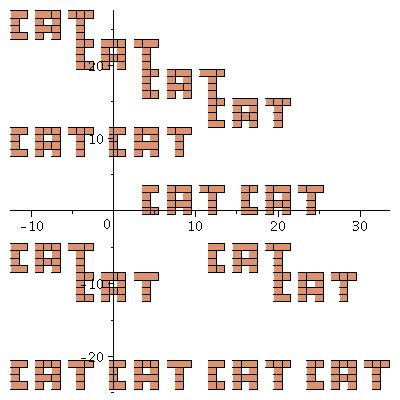}}
\put(-20,-5){$V^{[11]}$}\put(120,-5){$V^{[12]}$}
\end{picture}
\caption{Example~\ref{3-d-cat}: States on step 11 (left) and 12 (right).}\label{quat3}
\end{figure}

\begin{ex}\label{3-d-cat}{\rm
In this example $A$ is a free abelian group with two generator and $G$ - the group of quaternions of order 8, so, modulus $p$ is equal to $2$. The number $\kappa$ for $G$ is equal to 2 because 4 is the maximal order of elements in quaternions group. 

Fig.~\ref{quat1} represents initial state $V$ with $\proj_G(V)=\{\m1_G\}$ (so, it is two-dimentional in a clear sense) and rule $R$ with $\supp(R)=\{(-1,-1,2),(1,-2,2),(-2,2,1),(1,2,3)\}$. Projections of $\supp(R)$ on both factors $A,G$ of the product $G\times A$ are not trivial. 

Accordingly to corollary~\ref{vanishing} on steps whose numbers are multiple 4(=$p^{\kappa}$) states must be empty or project into $\{\m1_G\}$. 

Fig.~\ref{quat2} represents states on steps 1 and 4 whereas states on steps 11 and 12 are shown on Fig.~\ref{quat3}. We can see that states on steps 4 and 12 are ``flat''.
}\bx
\end{ex}

\begin{figure}[here]
 \begin{picture}(100,150)(0,0)
\put(-60,0){\includegraphics[width=4.5cm]{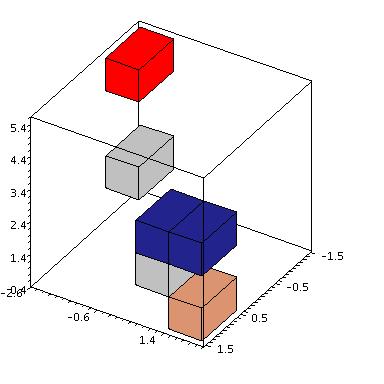}}
\put(60,16){\includegraphics[width=6cm]{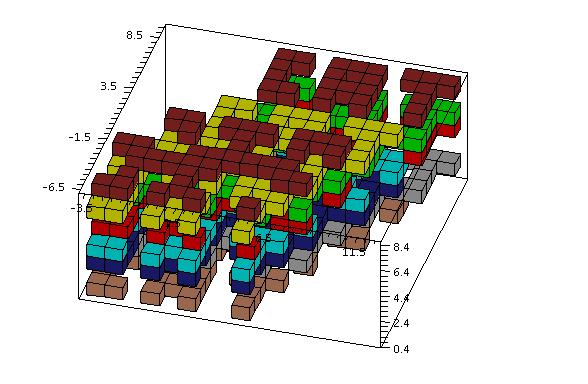}}
\put(-20,-5){$R$}\put(120,-5){$V^{[3]}$}
\end{picture}
\caption{Example~\ref{3-d-cat-empty}: Rule (left) and state on step 3 (right).}\label{empty}
\end{figure}
Whether states project into $\{\m1_G\}$ at moments multiple to $2^{\kappa}$  or become empty starting from $t=2^{\kappa}$, it depend on the given rule. 

\begin{ex}\label{3-d-cat-empty}{\rm
Here $p=2$, $G$ is quaternion group of order 8, and the initial state $V$ is the same as in example~\ref{3-d-cat}, whereas rule $R$ is shown on the left part of Fig.~\ref{empty}. $\supp(R)=\{(1,2,1),(1,1,2),(1,1,3),91,2,3),(-1,-2,5)\}$. The right part of Fig.~\ref{empty} shows $V^{[3]}$. The state $V^{[4]}$ and all next states appear to be empty.
}\bx
\end{ex}

\begin{figure}[here]
 \begin{picture}(100,150)(0,0)
\put(-110,16){\includegraphics[width=3.5cm]{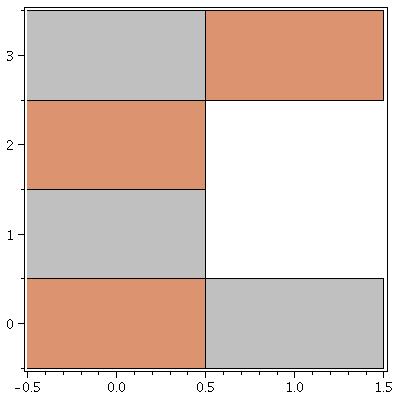}}
\put(30,6){\includegraphics[width=4.5cm]{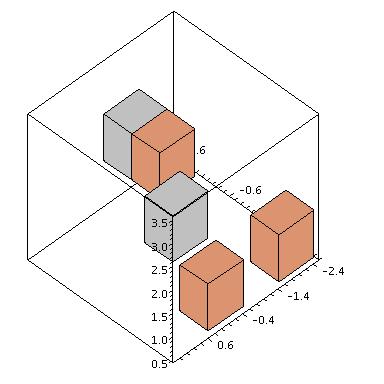}}
\put(-60,-5){$V$}\put(60,-5){$R$}
\end{picture}
\caption{Group $G$ (see Remark) of order 27, modulus $p=3$: 
Initial state $V$ and rule $R$. Brown color corresponds value 1 of a cell, gray - value 2.}\label{G27a}
\end{figure}
\begin{figure}[here]
 \begin{picture}(100,150)(0,0)
\put(-90,10){\includegraphics[width=5.0cm]{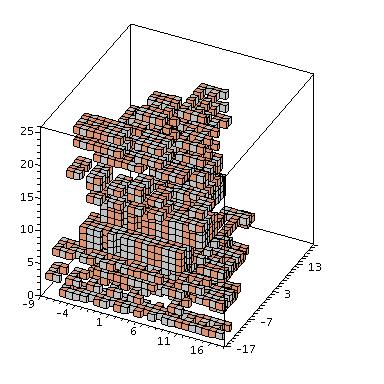}}
\put(70,16){\includegraphics[width=4.0cm]{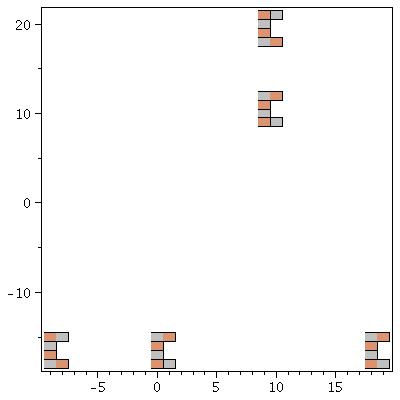}}
\put(-20,-5){$V^{[8]}$}\put(120,-5){$V^{[9]}$}
\end{picture}
\caption{Group $G$ (see Remark) of order 27, modulus $p=3$: states on step 8 (left) and 9 (right).}\label{G27b}
\end{figure}
It appears that ACA on groups $G\times A$ (where $A$ is abelian) demonstrate similar behavior also for some non-commutative $p$-groups $G$ not included into the class $\mathrm{DHC}(\mbf S,\theta)$. 
\begin{ex}\label{withG27}{\rm
Let $p=3$. Fig.~\ref{G27a}, Fig.~\ref{G27b}  
illustrate the case when $G$ is a noncommutative groups of order 27 with the identity $x^3=1$ (Burnside $p^3$-group, see \cite{Burnside}, p.145, case II(v). A table for the group see Appendix). The description of rule $R$ as a vector is the following:
\beas
R(-1,-1,2)=R(-2,2,1)=R(0,2,1)=1\\
R(1,2,3)=R(-1,-2,2)=2.
\eeas}\bx
\end{ex}
Thus the results above could be extended to some other modules $p$ and direct products of abelian groups with some other non-commutative $p$-groups.

\section{Some other non-commutative groups: glider guns}

ACA on non-commutative index groups could demonstrate behaviors very different from the behaviors of ACA on abelian groups. Some simple and bright examples remind glider guns from the scope of behavious of the automaton {\sl Life} of J.Conway. 
\begin{figure}[here]
 \begin{picture}(100,150)(0,0)
\put(-110,16){\includegraphics[width=2.0cm]{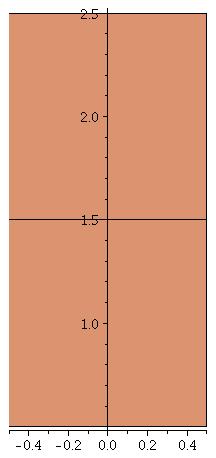}}
\put(30,6){\includegraphics[width=4.0cm]{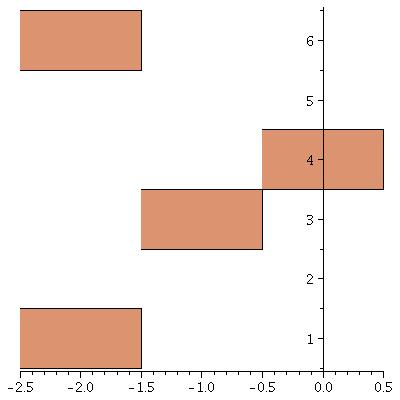}}
\put(-90,-5){$I$}\put(90,-5){$R$}
\end{picture}
\caption{Group $\mathrm{DIH3}\times\mathbb Z$, modulus $p=2$: 
Initial state $I$ and rule $R$.}\label{gun-R-I}
\end{figure}

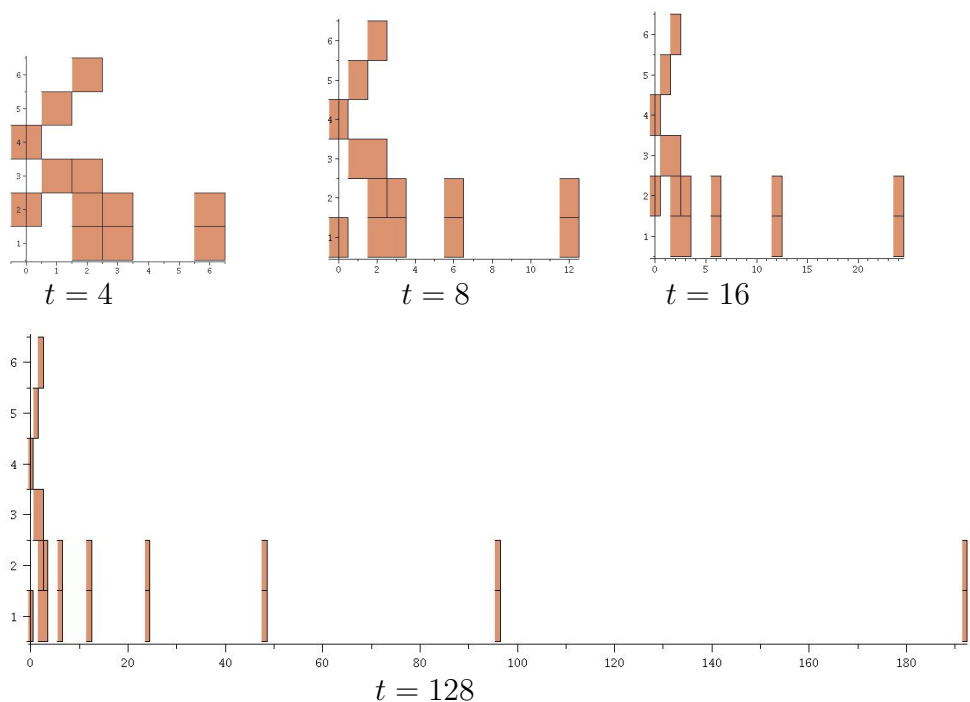
\begin{figure}[here]
 \begin{picture}(100,280)(0,0)
\put(-120,160){\includegraphics[width=3cm]{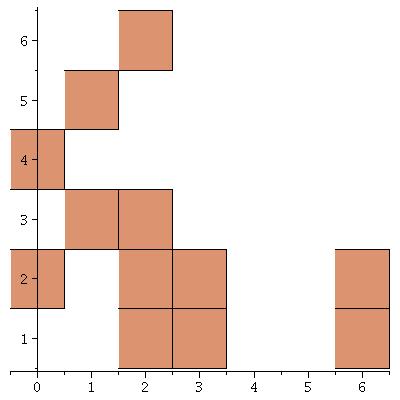}}
\put(0,160){\includegraphics[width=3.5cm]{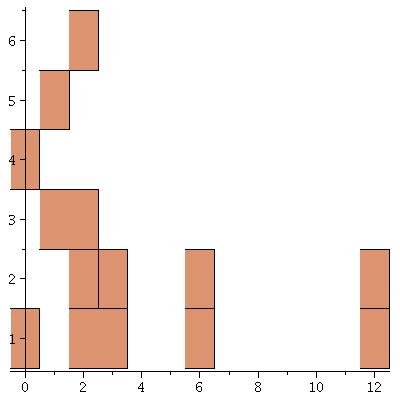}}
\put(120,160){\includegraphics[width=3.6cm]{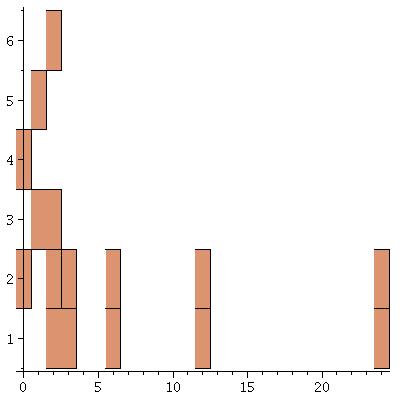}}
\put(-105,150){$t=4$}\put(30,150){$t=8$}\put(130,150){$t=16$}

\put(-120,10){\includegraphics[width=13cm]{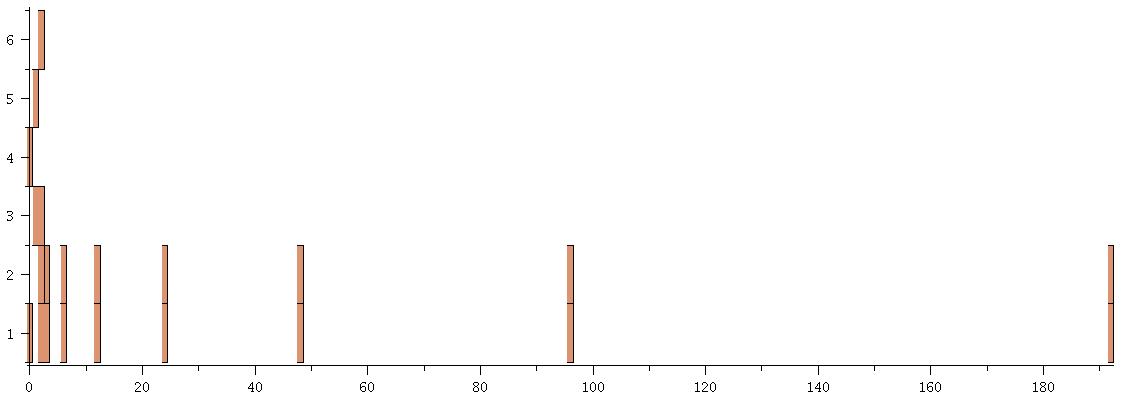}}
\put(20,0){$t=128$}
\end{picture}
\caption{States $I_0^{[t]}$ at time moments 4,8,16,128.}\label{I4-I128}
\end{figure}

Here we consider an example for a ACA on index group $\mathrm{DIH3}\times\mathbb Z$ over field $\m F_2$.  
It is convenient to visualize automata states. For that we represent the group $\mathrm{DIH3}$ (see \cite{Dih1}) by elements $1,2,3,4,5,6$ 
and multiplication table
\beas 
\left[ \begin {array}{cccccc} 1&2&3&4&5&6\\\noalign{\medskip}2&4&5&1&
6&3\\\noalign{\medskip}3&6&1&5&4&2\\\noalign{\medskip}4&1&6&2&3&5
\\\noalign{\medskip}5&3&2&6&1&4\\\noalign{\medskip}6&5&4&3&2&1
\end {array} \right] 
\eeas

\begin{figure}[here]
 \begin{picture}(100,110)(0,0)
\put(-90,6){\includegraphics[width=3.5cm]{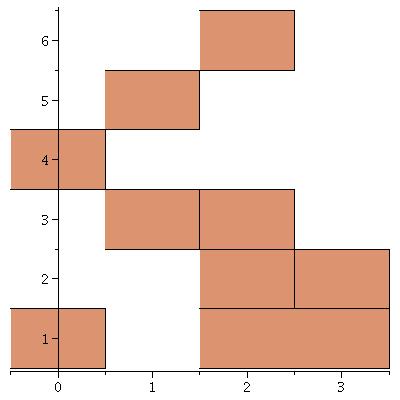}}
\put(60,6){\includegraphics[width=3.5cm]{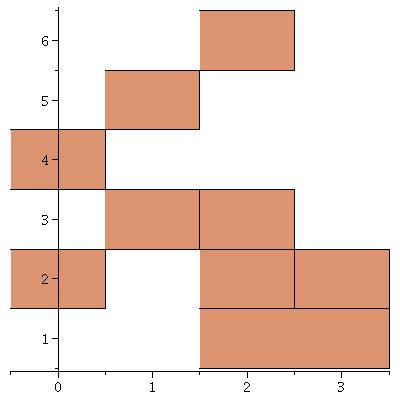}}
\put(-60,-5){$B$}\put(90,-5){$C$}
\end{picture}
\caption{States $B$ and $C$.}\label{B-C}
\end{figure}

The whole group $\mathrm{DIH3}\times\mathbb Z$ can be represented as a band $\{(x,y)\text{\bf: } x\in\{1,2,3,4,5,6\}, y\in\mathbb Z\}$ on Eucledian plane with component-wise group operation on pairs. 

Our ACA is defined by rule $R$ and initial state $I$ (shown on Fig.~\ref{gun-R-I}). States $I^{[t]}$ at time moments $t=8,16,128$ are shown by Fig.~\ref{I4-I128}. We see that $I^{[t]}$ for these time moments are built of subvectors (configurations) of three sorts $I,B,C$. (Configurations $B,C$ are shown on Fig.~\ref{B-C}.) The configurations can occupy different positions, and we record a position of a configuration $X\in\{I,B,C\}$ by integer $\min(\proj_{\mathbb Z}(X))$. That is $X_j$ denote a configuration of kind $X$ occupying a position in the band such that the least number of its projection on the group factor $\mathbb Z$ is equal to $j\in\mathbb Z$. In particular $I=I_0$.

In these terms Fig.~\ref{gun-R-I} (left side) shows $I_0$ and Fig.~\ref{B-C} shows $B_0,C_0$. We have
\bea
\supp(I_0)=\{(0,1),(0,2)\},\label{suppI}\\
\quad\supp(B_0)=\{(0,1),(0,4),(1,3),(1,5),(2,1),(2,2),(2,3),(2,6),(3,1),(3,2)\},\label{suppB}\\
\quad\supp(C_0)=\{(0,2),(0,4),(1,3),(1,5),(2,1),(2,2),(2,3),(2,6),(3,1),(3,2)\}.\label{suppC}
\eea  
Due to $p=2$ for any $z\in\mathbb Z$ it holds:
\bea
I_z+B_z+C_z=0.
\eea

\begin{figure}[here]
 \begin{picture}(100,110)(0,0)
\put(-90,6){\includegraphics[width=3.5cm]{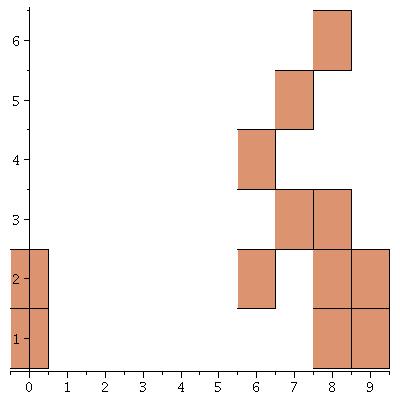}}
\put(60,6){\includegraphics[width=3.5cm]{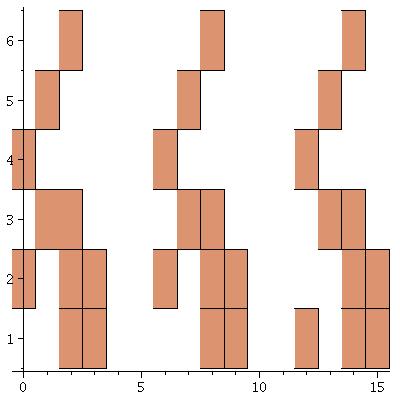}}
\put(-60,-5){$B_4$}\put(90,-5){$B_8$}
\end{picture}
\caption{States $B_4$ and $B_8$.}\label{B_4-B_8}
\end{figure}

\begin{figure}[here]
 \begin{picture}(100,110)(0,0)
\put(-90,6){\includegraphics[width=3.5cm]{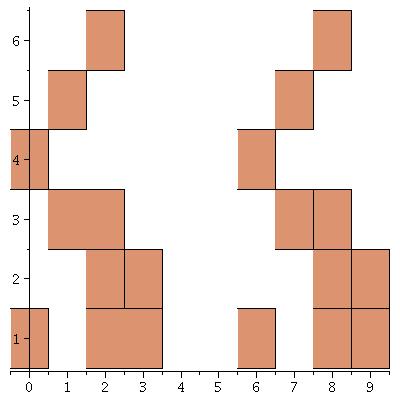}}
\put(60,6){\includegraphics[width=3.5cm]{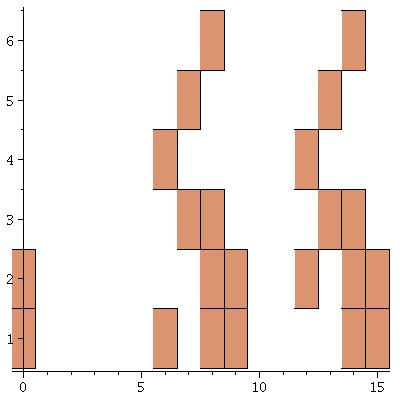}}
\put(-60,-5){$C_4$}\put(90,-5){$C_8$}
\end{picture}
\caption{States $C_4$ and $C_8$.}\label{C_4-C_8}
\end{figure}

Taking into account that as vectors all $B_i,C_j,I_k,i,j,k\in\mathbb Z,$ are defined on complete group $\mathrm{DIH3}\times\mathbb Z$, we can describe some states of automata in terms of sum by modulo $p,p=2,$ vectors $B_i,C_j,I_k$.  For example, $I^{[8]},I^{[16]},I^{[128]}$ have the following representations:
\beas
I^{[8]}=B_0+I_6+I_{12},\\
I^{[16]}=C_0+I_6+I_{12}+I_{24},\\
I^{[128]}=B_0+I_6+I_{12}+I_{24}+I_{48}+I_{96}+I_{192}.
\eeas  
\begin{theo}\label{g-gun}
For additive ACA on the group $\mathrm{DIH3}\times\mathbb Z$ with initial state $I_0$ and rule $R$ it holds
\bea\label{I2k}
I_0^{[2^k]}=
\begin{cases}
B_0+\sum_{i=2}^kI_{6\cdot2^{i-2}},& \text{if }k\text{ is odd},\\ 
C_0+\sum_{i=2}^kI_{6\cdot2^{i-2}},& \text{if }k\text{ even},\\
\end{cases}
\eea
for all $k\in\mathbb Z, k\ge2.$  
\end{theo}
\pr 
For configuration $X$ let $s^k(V)$ denote a shift of it on $k$ positions right along the axis $\mathbb Z$. That is $z\in\supp(\proj_{\mathbb Z}(s^k(X)))\iff z+k\in \supp(\proj_{\mathbb Z}(X))$. Then by the definition of ACA we have $R*(s(X))=s(R*X)$, in other words operations $R*\cdot$ and $s(\cdot )$ commute. 

Now we formulate two more statements ($k\ge 2$):
\bea\label{B2k}
B_0^{[2^k]}=
\begin{cases}
C_0+C_6+\sum_{i=3}^kB_{6\cdot2^{i-2}},& \text{if }k\text{ is odd},\\ 
I_0+C_6+\sum_{i=3}^kI_{6\cdot2^{i-2}},& \text{if }k\text{ even},\\
\end{cases}
\eea 
and 
\bea\label{C2k}
C_0^{[2^k]}=
\begin{cases}
I_0+B_6+\sum_{i=3}^kC_{6\cdot2^{i-2}},& \text{if }k\text{ is odd},\\ 
B_0+B_6+\sum_{i=3}^kI_{6\cdot2^{i-2}},& \text{if }k\text{ even},\\
\end{cases}
\eea 
and prove (\ref{I2k}), (\ref{B2k}), (\ref{C2k}) by a parallel induction on $k$.

{\sl Basis: $k=2,3$.} Correctness of the statements~(\ref{B2k}) and (\ref{C2k}) has been demonstrated by direct calculations whose results are presented on Figures~\ref{B_4-B_8}, \ref{C_4-C_8}.

{\sl Inductive step: $k\to k+1$.} We start with $I_0^{[2^{k+1}]}$ and odd $k$. Note, that by the definition $X^{[i]}$ is the state  at time $i$ of a ACA with an initial state $X$. Therefore $X^{[k+j]}= \left( X^{[k]} \right)^{[j]}$ for any $k,j\in\mathbb Z^+$. Thus
\beas
I_0^{[2^{k+1}]}=\left(I_0^{[2^k]}\right)^{[2^k]}= \left(B_0+\sum_{i=2}^kI_{6\cdot2^{i-2}}\right)^{[2^k]}=
B_0^{[2^k]}+\sum_{i=2}^kI^{[2^k]}_{6\cdot2^{i-2}}=\\
(\text{at this point we use hypothesis for $B_0^{[2^k]}$ and $I^{[2^k]}_{6\cdot2^{i-2}}$})\\
=C_0+C_6+\sum_{i=3}^kB_{6\cdot2^{i-2}}+ \sum_{i=2}^k\left(B_{6\cdot2^{i-2}} +\sum_{j=2}^kI_{6\cdot2^{i-2}+6\cdot2^{j-2}}\right)=\\
(\text{we cancell equal terms by modulo 2. In particular, terms $I_{6\cdot2^{i-2}+{6\cdot2^{j-2}}}$ with $i\neq j$ disappear})\\
=C_0+C_6+B_6+\sum_{i=2}^kI_{6\cdot2^{i-1}}= C_0+I_6+\sum_{i=3}^{k+1}I_{6\cdot2^{i-2}}=\\
C_0+\sum_{i=2}^{k+1}I_{6\cdot2^{i-2}}.
\eeas 
Similarly for even $k$ we obtain
\beas
I_0^{[2^{k+1}]}= \left(C_0+\sum_{i=2}^kI_{6\cdot2^{i-2}}\right)^{[2^k]}=
B_0^{[2^k]}+\sum_{i=2}^kI^{[2^k]}_{6\cdot2^{i-2}}=\\
(\text{at this point we use the hypothesis for $C_0^{[2^k]}$ and $I^{[2^k]}_{6\cdot2^{i-2}}$})\\
=B_0+B_6+\sum_{i=3}^kC_{6\cdot2^{i-2}}+ \sum_{i=2}^k\left(C_{6\cdot2^{i-2}} +\sum_{j=2}^kI_{6\cdot2^{i-2}+6\cdot2^{j-2}}\right)=\\
=B_0+B_6+C_6+\sum_{i=2}^kI_{6\cdot2^{i-1}}= B_0+I_6+\sum_{i=3}^{k+1}I_{6\cdot2^{i-2}}=\\
B_0+\sum_{i=2}^{k+1}I_{6\cdot2^{i-2}}.
\eeas 
Now we consider $B_0^{[2^{k+1}]}$ for odd $k$:
\beas
B_0^{[2^{k+1}]}=\left(C_0+C_6+\sum_{i=3}^kB_{6\cdot2^{i-2}}\right)^{[2^k]} =C_0^{[2^k]}+C_6^{[2^k]}+\sum_{i=3}^kB_{6\cdot2^{i-2}}^{[2^k]}=\\
\left\{I_0+B_6+ \sum_{i=3}^kC_{6\cdot2^{i-2}}\right\}+ \left\{I_6+B_{12}+ \sum_{i=3}^kC_{6+6\cdot2^{i-2}}\right\}+ 
\sum_{i=3}^k\Big(C_{6\cdot2^{i-2}}+C_{6+6\cdot2^{i-2}}+\\ \sum_{j=3}^kB_{6\cdot2^{i-2}+6\cdot2^{j-2}}\Big)= I_0+I_6+B_6+B_{12}+ 
\sum_{i=3}^kB_{6\cdot2^{i-1}}=\\
I_0+C_6+B_{12}+\sum_{i=4}^{k+1}kB_{6\cdot2^{i-2}}=I_0+C_6+\sum_{i=3}^{k+1}kB_{6\cdot2^{i-2}}
\eeas  
And for even $k$:
\beas
B_0^{[2^{k+1}]}=\left(I_0+C_6+\sum_{i=3}^kB_{6\cdot2^{i-2}}\right)^{[2^k]} =I_0^{[2^k]}+C_6^{[2^k]}+\sum_{i=3}^kB_{6\cdot2^{i-2}}^{[2^k]}=\\
\left\{C_0+\sum_{i=3}^kI_{6\cdot2^{i-2}}\right\}+ \left\{B_6+B_{12}+ \sum_{i=3}^kC_{6+6\cdot2^{i-2}}\right\}+ 
\sum_{i=3}^k\Big(I_{6\cdot2^{i-2}}+C_{6+6\cdot2^{i-2}}+\\ \sum_{j=3}^kB_{6\cdot2^{i-2}+6\cdot2^{j-2}}\Big)= C_0+I_6+B_6+B_{12}+ 
\sum_{i=3}^kB_{6\cdot2^{i-1}}=\\
C_0+C_6+B_{12}+\sum_{i=4}^{k+1}kB_{6\cdot2^{i-2}}=C_0+C_6+\sum_{i=3}^{k+1}kB_{6\cdot2^{i-2}}.
\eeas  
It is possible to get the expression for $C^{[2^{k+1}}$ directly or by the symmetry
\beas
(k\text{ odd },B,C)\iff (k\text{ even },C,B)
\eeas 
which holds for equations~(\ref{B2k}), (\ref{C2k}). That is if we replace words ``odd'', with ``even'' and vice versa, and all occurences of $B$ ($C$) replace with $C$ ($B$) respectively, the system does not change. With this symmetry the induction step for $C$ can be obtained from the inductive step for $B$. \bx

Despite the nature of the described behavior differs from the case of the automaton {\sl Life}, additive glider guns for non-commutative groups can be viewed as analogues of glider guns from {\sl Life}. Also as it follows from the results for commutative groups, no additive automata exist on commutative index-groups that are able to demonstrate the behaviour described in theorem~\ref{g-gun}. On the other hand, any commutative group can be extended in a way to a non-commutative group. Therefore obviously for any additive CA $A$ on abelian group there exists an appropriate non-commutative group and suitable additive CA on it that represents the behavior of $A$. 

\begin{cor}
Class behavior patterns of additive CA on non-commutative groups is essentially wider the class of behavior patterns of additive CA on commutative groups.
\end{cor}

\section{Conclusion} 

Thus for ACA on groups, behavior can simplify essentially if observed in time moments from a special infinite series. Moreover, the size of observed parts of finite direct factors and even the observed topology (opportunity to notice some dimentions) can depend on the length of time step of observations. 

As we noted above some typical behaviors of ACA on commutative groups in special time moments could be viewed as gliders' flights from the well known automaton {\sl Life} of J. Conway. Another pattern of behavior for the latter automaton is gilder gun. These two phenomena play important role in the theory of the automaton {\sl Life}. Additive cellular automata on some non-commutative groups can demonstrate behaviors simulating gliders and glider guns.

\newpage

\appendix
\section*{Appendix}

This is the table for the group $G$ from example~\ref{withG27}.

\tiny

\beas
\hspace{-1cm} \left[ \begin {array}{ccccccccccccccccccccccccccc} 1&2&3&4&5&6&7&8&9&
10&11&12&13&14&15&16&17&18&19&20&21&22&23&24&25&26&27
\\\noalign{\medskip}2&4&5&1&9&8&11&16&3&14&17&18&15&20&24&6&7&22&25&10
&23&12&26&13&27&21&19\\\noalign{\medskip}3&6&7&10&13&12&1&17&18&15&21&
2&19&22&4&24&25&23&5&11&20&27&9&26&8&16&14\\\noalign{\medskip}4&1&9&2&
3&16&17&6&5&20&7&22&24&10&13&8&11&12&27&14&26&18&21&15&19&23&25
\\\noalign{\medskip}5&8&11&14&15&18&2&7&22&24&23&4&25&12&1&13&27&26&9&
17&10&19&3&21&16&6&20\\\noalign{\medskip}6&10&13&3&18&17&21&24&7&22&25
&23&4&11&26&12&1&27&8&15&9&2&16&19&14&20&5\\\noalign{\medskip}7&12&1&
15&19&2&3&25&23&4&20&6&5&27&10&26&8&9&13&21&11&14&18&16&17&24&22
\\\noalign{\medskip}8&14&15&5&22&7&23&13&11&12&27&26&1&17&21&18&2&19&
16&24&3&4&6&25&20&10&9\\\noalign{\medskip}9&16&17&20&24&22&4&11&12&13&
26&1&27&18&2&15&19&21&3&7&14&25&5&23&6&8&10\\\noalign{\medskip}10&3&18
&6&7&24&25&12&13&11&1&27&26&15&19&17&21&2&14&22&16&23&20&4&5&9&8
\\\noalign{\medskip}11&18&2&24&25&4&5&27&26&1&10&8&9&19&14&21&16&3&15&
23&17&20&22&6&7&13&12\\\noalign{\medskip}12&15&19&7&23&25&20&26&1&27&8
&9&10&21&16&2&3&14&17&4&18&6&24&5&22&11&13\\\noalign{\medskip}13&17&21
&22&4&23&6&1&27&26&9&10&8&2&3&19&14&16&18&25&15&5&7&20&24&12&11
\\\noalign{\medskip}14&5&22&8&11&13&27&18&15&17&2&19&21&24&25&7&23&4&
20&12&6&26&10&1&9&3&16\\\noalign{\medskip}15&7&23&12&1&26&8&2&19&21&3&
14&16&4&5&25&20&6&22&27&24&9&11&10&13&18&17\\\noalign{\medskip}16&20&
24&9&12&11&26&15&17&18&19&21&2&7&23&22&4&25&6&13&5&1&8&27&10&14&3
\\\noalign{\medskip}17&22&4&13&27&1&9&19&21&2&14&16&3&25&20&23&6&5&24&
26&7&10&12&8&11&15&18\\\noalign{\medskip}18&24&25&11&26&27&10&21&2&19&
16&3&14&23&6&4&5&20&7&1&22&8&13&9&12&17&15\\\noalign{\medskip}19&25&20
&27&10&9&12&3&14&16&18&15&17&6&7&5&22&24&23&8&4&13&1&11&26&2&21
\\\noalign{\medskip}20&9&12&16&17&15&19&22&24&7&4&25&23&13&27&11&26&1&
10&18&8&21&14&2&3&5&6\\\noalign{\medskip}21&23&6&26&8&10&13&14&16&3&15
&17&18&5&22&20&24&7&4&9&25&11&27&12&1&19&2\\\noalign{\medskip}22&13&27
&17&21&19&14&23&4&25&6&5&20&26&8&1&9&10&11&2&12&16&15&3&18&7&24
\\\noalign{\medskip}23&26&8&21&16&14&15&20&6&5&24&7&22&9&12&10&13&11&1
&3&27&17&19&18&2&25&4\\\noalign{\medskip}24&11&26&18&2&21&16&4&25&23&5
&20&6&1&9&27&10&8&12&19&13&3&17&14&15&22&7\\\noalign{\medskip}25&27&10
&19&14&3&18&5&20&6&22&24&7&8&11&9&12&13&26&16&1&15&2&17&21&4&23
\\\noalign{\medskip}26&21&16&23&6&20&24&10&8&9&13&11&12&3&18&14&15&17&
2&5&19&7&25&22&4&27&1\\\noalign{\medskip}27&19&14&25&20&5&22&9&10&8&12
&13&11&16&17&3&18&15&21&6&2&24&4&7&23&1&26\end {array} \right] 
\eeas
\end{document}